\documentclass{article}
\usepackage[utf8]{inputenc}
\usepackage{graphics}
\usepackage{graphicx,amsmath,bm,amssymb,pdfpages}
\usepackage[none]{hyphenat}
\usepackage[square,numbers]{natbib}
\usepackage{xcolor}
\usepackage{soul}
\usepackage{authblk}
\usepackage{mathrsfs}
\usepackage{here}
\title{Exact Solutions Of Time Fractional Generalized  Burgers-Fisher Equation Using Exp function and Exponential Rational Function Methods}

\begin{document}
\author[1]{Ramya Selvaraj\thanks{ramya@src.sastra.edu}}
\author[2]{Swaminathan Venkatraman\thanks{Corresponding Author: swaminathan@src.sastra.edu}}
\affil[1]{Department of Mathematics, Srinivasa Ramanujan Centre, SASTRA Deemed to be University, Kumbakonam - 612 001, Tamil Nadu, India.}
\affil[2]{Discrete Mathematics Laboratory, Department of Mathematics, Srinivasa Ramanujan Centre, SASTRA Deemed to be University, Kumbakonam - 612 001, Tamil Nadu, India.}
\date{}

\maketitle
\abstract{
 Using modified Riemann-Liouville derivative, the Exp function and Exponential rational function methods are implemented to solve the time fractional generalized Burgers-Fisher equation (TF-GBF). The TF-GBF is transformed into a nonlinear ordinary differential equation (NLODE) by applying the transformation of traveling wave. The suggested methods are then introduced to formulate exact solutions for the resulting equation. The solutions are depicted using 2D and 3D plots.}
\section{Introduction}

In the year 1695, Leibnitz introduced a fractional calculus. Several studies have been carried out on science and engineering related to fractional phenomena such as plasma physics, viscoelastic materials, fluid flow, biology, economics, probability and statistics, polymers, optical fibers, etc. In recent years, fractional differential equations (FDEs) have gained a considerable popularity among scientists and engineers and finding solutions for nonlinear FDE (NLFDE) is an indefatigable research field  \cite{oldham,podlubny,kilbas,juma,eslami,huan}. \\

Numerous mathematical procedures have been developed and investigated to find the exact solutions of the NLFDE.
For instance, the sine-cosine method \cite{Bekir,Mirzazadeh}, the tanh method \cite{Wazwaz}, Adomian decomposition method \cite{elsayed}, the sub-equation method \cite{Alzaidy,Guo}, variational iteration method \cite{wu}, the first integral method \cite{Lu,bekirunsal} and so on \cite{Liuchen,pandir,Geometrical}. Recently, an Exp-function method, which was proposed by He and Wu \cite{Hewu} and consistently studied in \cite{Noor,Ebaid,Heabu}. This method was previously suggested for evaluating a solution for PDEs. In addition, it was expanded successfully to FDEs \cite{expHE,gunerbekir,zhangzong,Bzheng,yan}.  \\

In addition, the exponential rational function method (ERFM) is one of the methods to reveal the exact solutions. At first, it was originated by \cite{demiray}. Furthermore, it has been executed in many fields of engineering and science \cite{aksoy,bekirkaplan,mohybibi}.
From the contribution of the prior mentioned schemes, we carry out this analysis to the time fractional generalized Burgers-Fisher equation (TF-GBFE).  A nonlinear equation is called the Burgers-Fisher equation and is the combination of reaction, convection and diffusion process. The properties of the convective phenomenon of Burgers and Fisher's diffusion transport as well as reaction form characteristics are used in the nonlinear equation. The GBFE is used in the field of fluid dynamics. It was also found in some applications including gas dynamics, heat conduction, elasticity, etc.\\

The GBFE with the order of time fractional $\alpha$ and an  arbitrary constants $\beta,\gamma$ and $\delta$
 is given by
\begin{eqnarray} \label{GBF}
u^\alpha _t+\beta u^\delta u_x-u_{xx}=\gamma u(1-u^\delta), 
\end{eqnarray}
where $\alpha \in (0,1] $.\\

The present work is classified as follows: Section 2 describes the properties of the modified Riemann-Liouville fractional derivative and the description and implementation of Exp function method in TF-GBFE is given in section 3. The description and implementation of ERFM in TF-GBFE is clarified in section 4.  Section 5 presents the results and discussion of the proposed method for TF-GBFE and section 6 ends with conclusion.

\section{The modified Riemann-Liouville (RL) derivative}
Recent studies have shown that the dynamics of many physical processes are precisely described using FDEs with various types of fractional derivatives. Jumarie has given a different interpretation of the fractional derivative with a minor modification of the  Riemann-Liouville (RL) derivative  \cite{Jumarie1,Jumarie2} and it is defined as
\begin{equation}
  D^\alpha_t g(t)=\left\{\begin{array}{ll}
\displaystyle{\frac{1}{\Gamma (1-\alpha)}\frac{d}{dt}\int_0^t(t-\zeta)^{-\alpha-1} [g(\zeta)-g(0)]d\zeta}, \;  0 < \alpha < 1, \\
\displaystyle{(g^{(\vartheta)}(t))^{(\alpha-\vartheta)}},  \vartheta \leq \alpha < \vartheta+1, \;\vartheta \geq 1, \end{array} \right.
\end{equation}
where $\alpha$ denotes the order of fractional derivative and $g(t)$ denotes a continuous non differentiable function $g:\mathbb{R} \to \mathbb{R}$, $t \to g(t)$. Some notable properties of modified RL derivative are listed below \cite{Heli,LiandHe}:\\
Property: 1\\

\[D^\alpha_t t^\kappa=\frac{\Gamma(1+\kappa)}{\Gamma(1+\kappa-\alpha)} t^{\kappa-\alpha},\; \kappa > 0.\]    

Property: 2\\

\[D^\alpha_t (p g(t)+q h(t))=p D^\alpha_t g(t)+q D^\alpha_t h(t).\]    
where $g$ and $h$ are constants.\\
Property: 3\\

\[D^\alpha_t k=0,\;k=constant.\]

For the following problem, these properties can be used.
\section{The Exp Function Method}
Let us consider the NLFDE with an unknown function $u$, the polynomial G of $u$ and its partial derivatives, involving the highest order derivatives and the nonlinear terms,
\begin{eqnarray} \label{FPDE}
  G(u, D^\alpha_tu, D^\beta_xu, D^\alpha_tD^\alpha_tu, D^\alpha_tD^\beta_xu, D^\beta_xD^\beta_xu,\ldots)=0,\; \; 0<\alpha, \beta<1.
\end{eqnarray}
With the following fractional complex transform \cite{LiandHe} with nonzero arbitrary constants $\lambda$ and $\xi$, we transform  fractional differential equation into ordinary differential equation:
    \begin{eqnarray}\label{travelingwave}
    u(x,t)=u(\zeta),    \;   \zeta=\frac{\lambda t^\alpha}{\Gamma[1+\alpha]}+\frac{\xi x^\beta}{\Gamma[1+\beta]}, 
\end{eqnarray}
Reduce Eq.(\ref{FPDE}) to the following ordinary differential equation (ODE) of integer order:
\begin{eqnarray} \label{conversionODE}
    H(u,u^\prime,u^{\prime \prime},u^{\prime \prime \prime}, \ldots)=0,
\end{eqnarray}
where prime denotes a derivative of $\zeta$.\\
 With the positive integers $g, h, k,$ and $j$, also with the unknown constants $a_q$ and $b_p$, the Exp function can be expressed in the form:
 \begin{eqnarray} \label{expfunction formula}
u(\zeta)=\frac{\sum^h_{q=-g}a_q e^{q\zeta}}{\sum^j_{p=-k}b_p e^{p\zeta}}.
\end{eqnarray}
We may evaluate the values of $h, j$ and $g, k$ respectively by balancing the highest order and the lowest order within the Exp function.

\subsection{Exp function for the time fractional generalized Burgers-Fisher equation }
We apply the following transformation in order to solve the TF-GBFE using the proposed method:
\begin{eqnarray}\label{expfuntravelingwave}
    u(x,t)=u(\zeta),    \;   \zeta=kx-\frac{\lambda t^\alpha}{\Gamma[1+\alpha]}
\end{eqnarray}
Now Eq.(\ref{expfuntravelingwave}) in Eq.(\ref{GBF}), we get
  \begin{eqnarray} \label{tvwreductioneqn}
        k^2 u^{\prime\prime}+(\lambda-k\beta u^{\delta})u^{\prime}+\gamma u(1-u^{\delta})=0.
    \end{eqnarray}
Applying the following folding transformation in Eq.(\ref{tvwreductioneqn})
    \begin{eqnarray}
        u(\zeta)=v^{\frac{1}{\delta}}(\zeta),
    \end{eqnarray}
we obtain the following NLODE:
 \begin{equation} \label{nonlinearode}
   k^2\delta vv^{\prime \prime}+ k^2(1-\delta){v^\prime}^2+( \lambda -k \beta v)\delta v {v^\prime} +\gamma \delta^2 (1-v)v^2=0.
\end{equation}
According to Exp function method, we assume that the solution of Eq.(\ref{nonlinearode}) can be expressed in the form
\begin{eqnarray} \label{expfunction formula in GBF}
u(\zeta)=\frac{\sum^h_{q=-g}a_q e^{q\zeta}}{\sum^j_{p=-k}b_p e^{p\zeta}} = \frac{a_{-g}e^{-g\zeta}+\dots+a_{k}e^{k\zeta}}{b_{-h}e^{-h\zeta}+\dots+b_{j}e^{j\zeta}}.
\end{eqnarray}
Balancing the highest order in Eq.(\ref{nonlinearode}) for  $ vv^{\prime \prime}$ and $ v^2v^\prime$, we have


\begin{equation} \label{vvdoublehighest}
    vv^{\prime \prime}=\frac{h_5e^{(2k+3j)\zeta+\dots}}{h_6e^{(5j)\zeta+ \dots}}
\end{equation}
and
\begin{equation} \label{vsquarevhighest}
    v^2 v^{\prime}=\frac{h_9e^{(3k+j)\zeta}+\dots}{h_{10} e^{(4j)\zeta}+\dots}=\frac{h_9e^{(3k+2j)\zeta}+\dots}{h_{10} e^{(5j)\zeta}+\dots}.
\end{equation}
Balancing highest order of Exp function in Eq.(\ref{vvdoublehighest}) and Eq.(\ref{vsquarevhighest}), we obtain

\begin{equation}
    2k+3j=3k+2j , \;\; k=j.
\end{equation}
Likewise, balancing the lowest order in Eq.(\ref{nonlinearode}) for $ vv^{\prime \prime}$ and $ v^2v^\prime$, we have
\begin{equation} \label{vvdoublelowest}
    vv^{\prime \prime}=\frac{a_5e^{-(2g+3h)\zeta}+\dots}{a_6e^{(-5h)\zeta}+\dots}
\end{equation}
and
\begin{equation} \label{vsquarevlowest}
    v^2 v^{\prime}=\frac{a_9e^{-(3g+h)\zeta}+\dots}{a_{10}e^{(-4h)\zeta}+\dots}=\frac{a_9e^{-(3g+2h)\zeta}+\dots}{a_{10}e^{(-5h)\zeta}+\dots}.
\end{equation}
Balancing lowest order of Exp function in Eq.(\ref{vvdoublelowest}) and Eq.(\ref{vsquarevlowest}), we obtain
\begin{equation}
    2g+3h=3g+2h , \;\;g=h.
\end{equation}
For the choices of $k=j=1$ and $g=h=1$, Eq.(\ref{expfunction formula in GBF}) becomes,
\begin{equation} \label{GBFexpfunctionequation}
    v(\zeta)= \frac{a_1 e^{(\zeta)}+a_0+a_{-1} e^{(-\zeta)}}{b_1 e^{(\zeta)}+b_0+b_{-1} e^{(-\zeta)}}.
\end{equation}
Substituting Eq.(\ref{GBFexpfunctionequation}) into Eq.(\ref{nonlinearode}), we obtain the following cases with different parametric choices of  $a_{-1},a_{0},a_{1},b_{-1},b_{0}$ and $b_{1}$.\\
Case:1\\
 \[a_{-1}=0,\;a_0=0,\;b_{-1}=0,\;b_0=0,\;\lambda=\frac{-k^2-\gamma \delta^2}{\delta},k=\frac{-\gamma \delta}{\beta}, \]
Eq.(\ref{nonlinearode}) becomes
\begin{equation}
    -\gamma \delta v v^{\prime \prime}-\beta^2 v^2 (1+v^{\prime})+ \beta^2 v^3+ (\beta^2 +\gamma)v v^{\prime} +\gamma(\delta-1)=0,
\end{equation} 
we obtain 
\begin{equation} 
    v(x,t)=\frac{a_1 e^{\big({\frac{kx {\Gamma[1+\alpha]}-\lambda t^\alpha}{{\Gamma[1+\alpha]}}\big)}}}{b_0},
\end{equation}
then the exact solution is
\begin{equation} \label{expfuncase1}
u(x,t)=\displaystyle \left[\frac{a_1 e^{\big({\frac{kx {\Gamma[1+\alpha]}-\lambda t^\alpha}{{\Gamma[1+\alpha]}}\big)}}}{b_0}\right]^ \frac{1}{\delta}. 
\end{equation}
Case:2\\
\[a_{-1}=0,\;a_0=0,\;b_{-1}=0,\;a_1=b_1,\;\gamma=\frac{2k^2}{\delta^2},\; \lambda=\frac{-4k^2-\gamma \delta^2}{2\delta},\; k=\frac{-\beta \delta}{1+\delta}, \]
Eq.(\ref{nonlinearode}) becomes
\begin{equation}
    \delta v v^{\prime \prime}+(2+(1+\delta)v^{\prime})v^2-(\delta-1){v^\prime}^2-3vv^\prime-2v^3=0,
\end{equation}
we obtain 
\begin{equation}
    v(x,t)=\frac{a_1 e^{\big({\frac{kx {\Gamma[1+\alpha]}-\lambda t^\alpha}{{\Gamma[1+\alpha]}}\big)}}}{a_1 e^{\big({\frac{kx {\Gamma[1+\alpha]}-\lambda t^\alpha}{{\Gamma[1+\alpha]}}\big)}}+b_0},
\end{equation}
then the exact solution is
\begin{equation} \label{expfuncase2}
u(x,t)=\displaystyle \left[\frac{a_1 e^{\big({\frac{kx {\Gamma[1+\alpha]}-\lambda t^\alpha}{{\Gamma[1+\alpha]}}\big)}}}{a_1 e^{\big({\frac{kx {\Gamma[1+\alpha]}-\lambda t^\alpha}{{\Gamma[1+\alpha]}}\big)}}+b_0}\right]^ \frac{1}{\delta}. 
\end{equation}
Case:3\\
\[a_{-1}=0,\;a_1=0,\;b_{-1}=0,\;b_0=0,\; \lambda=\frac{\beta^2 \gamma \delta+\gamma^2 \delta}{\beta^2},\;k=\frac{\gamma \delta}{\beta},\]
Eq.(\ref{nonlinearode}) becomes
\begin{equation}
    \gamma^2 \delta^2 v v^{\prime \prime}-\gamma^2 \delta (\delta-1){v^\prime}^2+\beta^2(\lambda-\gamma\delta v)v v^\prime-\beta^2\gamma \delta (v-1)v^2=0,
\end{equation}
we have
\begin{equation} 
    v(x,t)=\frac{a_0 e^{-\big({\frac{kx {\Gamma[1+\alpha]}-\lambda t^\alpha}{{\Gamma[1+\alpha]}}\big)}}}{b_1},
\end{equation}
then the exact solution is
\begin{equation} \label{expfuncase3}
u(x,t)=\displaystyle \left[\frac{a_0 e^{-\big({\frac{kx {\Gamma[1+\alpha]}-\lambda t^\alpha}{{\Gamma[1+\alpha]}}\big)}}}{b_1}\right]^ \frac{1}{\delta}. 
\end{equation}
Case:4\\
\[a_{-1}=0,\;b_{-1}=0,\;b_0=0,\; b_1=a_1,\;\delta=1,\;\lambda=\frac{\gamma^2  \delta }{\beta^2},\;k=\frac{\gamma \delta}{\beta},\]
Eq.(\ref{nonlinearode}) becomes
\begin{equation}
    \gamma(v^{\prime \prime}+v^{\prime})-\beta^2(v^{\prime}-1)v-\beta^2v^2=0,
\end{equation}
we have
\begin{equation}
    v(x,t)=1+\frac{a_0 e^{-\big({\frac{kx {\Gamma[1+\alpha]}-\lambda t^\alpha}{{\Gamma[1+\alpha]}}\big)}}}{a_1},
\end{equation}
then the exact solution is
\begin{equation} \label{expfuncase4}
u(x,t)=\displaystyle \left[1+\frac{a_0 e^{-\big({\frac{kx {\Gamma[1+\alpha]}-\lambda t^\alpha}{{\Gamma[1+\alpha]}}\big)}}}{a_1}\right]^ \frac{1}{\delta}. 
\end{equation}
Case:5\\
\[a_{-1}=0,\;a_0=b_0,\;a_1=0,\;b_{-1}=0,\;\lambda=-k^2+k\beta+\gamma\delta,\;k=\frac{ \beta \delta}{1+\delta},\]
Eq.(\ref{nonlinearode}) becomes
\begin{equation}
    \beta^2 \delta vv^{\prime \prime}-\beta^2(\delta-1){v^\prime}^2+(1+\delta)(\gamma+\gamma \delta-\beta^2v^{\prime})v^2+(\beta^2+\gamma(1+\delta^2))vv^{\prime}-\gamma(1+\delta)^2v^3=0,
\end{equation}
we obtain
\begin{equation}
   v(x,t)=\frac{b_0}{b_0+b_1e^{\big({\frac{kx {\Gamma[1+\alpha]}-\lambda t^\alpha}{{\Gamma[1+\alpha]}}\big)}}},
\end{equation}
then the exact solution is
\begin{equation} \label{expfuncase5}
u(x,t)=\displaystyle \left[\frac{b_0}{b_0+b_1e^{\big({\frac{kx {\Gamma[1+\alpha]}-\lambda t^\alpha}{{\Gamma[1+\alpha]}}\big)}}}\right]^ \frac{1}{\delta}. 
\end{equation}
Case:6\\
\[a_{-1}=0,\; a_0=b_0,\; b_{-1}=0,\;b_{1}=0,\;\delta=1,\; \lambda=-k^2+k\beta+\gamma\delta,\;  k=-\frac{\gamma \delta}{\beta}, \] 
Eq.(\ref{nonlinearode}) becomes
\begin{equation}
    \gamma(v^{\prime \prime}-v^{\prime})+\beta^2(1+v^{\prime})v-\beta^2v^2=0,
\end{equation}
we acquire
\begin{equation} 
    v(x,t)=1+\frac{a_1 e^{\big({\frac{kx {\Gamma[1+\alpha]}-\lambda t^\alpha}{{\Gamma[1+\alpha]}}\big)}}}{b_0},
\end{equation}
then the exact solution is
\begin{equation} \label{expfuncase6}
u(x,t)=\displaystyle \left[1+\frac{a_1 e^{\big({\frac{kx {\Gamma[1+\alpha]}-\lambda t^\alpha}{{\Gamma[1+\alpha]}}\big)}}}{b_0}\right]^ \frac{1}{\delta}. 
\end{equation}
Case:7\\
   \[a_{-1}=0, \;a_0=0, \; b_0=0, \; b_1=0, \;\lambda=\frac{-4k^2 - \gamma \delta^2}{2\delta}, \;k=\frac{-\gamma \delta}{2\beta}\]
 Eq.(\ref{nonlinearode}) becomes 
\begin{equation}
   \gamma\delta v v^{\prime\prime}-\gamma(\delta-1){v^\prime}^2-2(\beta^2+\gamma)v v^\prime+2\beta^2(2+v^\prime)v^2-4\beta^2v^3=0,
\end{equation}
we attain
\begin{equation} 
    v(x,t)=\frac{a_1 e^{\big({\frac{2kx {\Gamma[1+\alpha]}-2\lambda t^\alpha}{{\Gamma[1+\alpha]}}\big)}}}{b_{-1}},
\end{equation}
then the exact solution is
\begin{equation} \label{expfuncase7}
u(x,t)=\displaystyle \left[\frac{a_1 e^{\big({\frac{2kx {\Gamma[1+\alpha]}-2\lambda t^\alpha}{{\Gamma[1+\alpha]}}\big)}}}{b_{-1}}\right]^ \frac{1}{\delta}. 
\end{equation}
Case:8\\
  \[a_{-1}=0,\;a_0=0,\; a_1=b_1,\; b_0=0, \; \lambda=\frac{-4k^2 - \gamma \delta^2}{2\delta}, \; \gamma=\frac{2k^2(2+\delta)}{\delta^2},\;k=-\frac{\beta \delta}{2(1+\delta)} \]
  Eq.(\ref{nonlinearode}) becomes 
\begin{equation}
  \delta v v^{\prime \prime}+2(2+\delta+(1+\delta)v^\prime)v^2-(\delta-1){v^\prime}^2-(4+\delta)v v^\prime -2(2+\delta)v^3=0,
\end{equation}
we get
\begin{equation} 
    v(x,t)=\frac{b_1 e^{\big({\frac{2kx {\Gamma[1+\alpha]}-2\lambda t^\alpha}{{\Gamma[1+\alpha]}}\big)}}}{b_1 e^{\big({\frac{2kx {\Gamma[1+\alpha]}-2\lambda t^\alpha}{{\Gamma[1+\alpha]}}\big)}}+b_{-1}},
\end{equation}
then the exact solution is
\begin{equation} \label{expfuncase8}
u(x,t)=\displaystyle \left[\frac{b_1 e^{\big({\frac{2kx {\Gamma[1+\alpha]}-2\lambda t^\alpha}{{\Gamma[1+\alpha]}}\big)}}}{b_1 e^{\big({\frac{2kx {\Gamma[1+\alpha]}-2\lambda t^\alpha}{{\Gamma[1+\alpha]}}\big)}}+b_{-1}}\right]^ \frac{1}{\delta}. 
\end{equation}
Case:9\\
  \[a_0=0,\; a_1=b_1,\;b_{-1}=0, \; b_0=0, \;\delta=1,\; \lambda=\frac{\gamma^2\delta}{2\beta^2}, \; k=\frac{\gamma \delta}{2\beta} \]
  Eq.(\ref{nonlinearode}) becomes 
\begin{equation}
  \gamma(v^{\prime \prime}+2v^{\prime})-2\beta^2(v^{\prime}-2)v-4\beta^2v^2=0,
\end{equation}
we have
\begin{equation} 
    v(x,t)=1+\frac{ a_{-1} e^{\big({\frac{2kx {\Gamma[1+\alpha]}-2\lambda t^\alpha}{{\Gamma[1+\alpha]}}\big)}}}{b_1},
\end{equation}
then the exact solution is
\begin{equation} \label{expfuncase9}
u(x,t)=\displaystyle \left[1+\frac{ a_{-1} e^{\big({\frac{2kx {\Gamma[1+\alpha]}-2\lambda t^\alpha}{{\Gamma[1+\alpha]}}\big)}}}{b_1}\right]^ \frac{1}{\delta}. 
\end{equation}
\section{The Exponential Rational Function Method (ERFM)}
We start by considering an NLFDE with the polynomial $G$ of $u$ and its partial derivatives, that includes the derivatives of the highest order and nonlinear terms

\begin{eqnarray} \label{FPDE1}
  \mathscr{B}(u, D^\alpha_tu, D^\beta_xu, D^\alpha_tD^\alpha_tu, D^\alpha_tD^\beta_xu, D^\beta_xD^\beta_xu,\ldots)=0,\; \; 0<\alpha, \beta<1.
\end{eqnarray}
where the role of $u$ is unknown.\\
He and Wu \cite{Hewu} suggested a fractional complex transformation to turn the fractional differential equation into ordinary differential equation. Using the following fractional complex transform with nonzero arbitrary constants $\lambda$ and $\xi$
    \begin{eqnarray}\label{travelingwaveexprat1}
    u(x,t)=u(\zeta),    \;   \zeta=\frac{\lambda t^\alpha}{\Gamma[1+\alpha]}+\frac{\xi x^\beta}{\Gamma[1+\beta]}, 
\end{eqnarray}
we reduce Eq.(\ref{FPDE}) to the following ordinary differential equation (ODE) of integer order:
\begin{eqnarray} \label{conversionODEexprat1}
    \mathscr{F}(u,u^\prime,u^{\prime \prime},u^{\prime \prime \prime}, \ldots)=0,
\end{eqnarray}
where prime is the derivative with respect to $\zeta$.\\
The Exp rational function can be expressed as
 \begin{eqnarray} \label{exprationalfunction formula1}
u(\zeta)=\sum^N_{q=0}\frac{a_q}{(1+e^{\zeta})^q}.
\end{eqnarray}
where constants $a_q (a_q \neq 0)$ to be determined later.  Using balancing principal, $N$ can be determined. \\
Replace the Eq.(\ref{exprationalfunction formula1}) in Eq.(\ref{conversionODEexprat1}) and compile all the coefficients with the same power of $e^{m\zeta}(m=1, 2,\dots,6)$, transformed into another $e^{m\zeta}$ coefficient on the left side of the ODE, together. The unknown parameters for $a_q$ can be achieved by setting  $e^{m\zeta}(m=1, 2,\dots,6)$ to zero  in the list of algebraic equations. To solve the system of equation, an exact solution is to be developed for non-linear ODE.

\subsection{ERFM for the time fractional generalized Burgers-Fisher equation }
We apply the following transformation in order to solve the TF-GBFE using the proposed method
\begin{eqnarray}\label{exprationalfuntravelingwave}
    u(x,t)=u(\zeta),    \;   \zeta=kx-\frac{\lambda t^\alpha}{\Gamma[1+\alpha]},
\end{eqnarray}
 in Eq.(\ref{GBF}), we get
  \begin{eqnarray} \label{tvwreductioneqnforexprat}
        k^2 u^{\prime\prime}+(\lambda-k\beta u^{\delta})u^{\prime}+\gamma u(1-u^{\delta})=0.
    \end{eqnarray}
The following folding transformation is implemented in Eq.(\ref{tvwreductioneqn})
    \begin{eqnarray}
        u(\zeta)=v^{\frac{1}{\delta}}(\zeta),
    \end{eqnarray}
We obtain NLODE as follows:
 \begin{equation} \label{nonlinearodeforexprat}
   k^2\delta vv^{\prime \prime}+ k^2(1-\delta){v^\prime}^2+( \lambda -k \beta v)\delta v {v^\prime} +\gamma \delta^2 (1-v)v^2=0
\end{equation}
Balancing $v v^{\prime \prime}$ and $v^3$, we obtain $N=2$.
In accordance with ERFM, we presume that the solution of  Eq.(\ref{nonlinearode}) can be formulated as 
\begin{equation}  \label{expratfunforGBF}
    v(\zeta)=a_0+\frac{a_1}{1+e^{\zeta}}+\frac{a_2}{(1+e^\zeta)^2}.
\end{equation}
We have the following cases when we replace Eq.(\ref{expratfunforGBF}) with Eq.(\ref{nonlinearode}) and collect all terms with the same power with $e^{m\zeta} (m=1, 2, \dots, 7)$, and then equalize the coefficients to zero.\\
Case: 1\\
\[a_0=1,\; a_1=0,\; a_2=-1,\; \beta=\frac{2k(\delta-1)}{\delta},\; \gamma=-\frac{6k^2}{\delta},\]\[\lambda=\frac{4k^2+2k\beta-\gamma\delta}{2}, \;\delta=1, \]
Eq.(\ref{nonlinearode}) turns into
\begin{equation}
    v^{\prime \prime}+5v^{\prime}+6v^2-6v=0,
\end{equation}
we obtain
\begin{equation}
    v(x,t)=\frac{2 e^{\big({\frac{kx {\Gamma[1+\alpha]}-\lambda t^\alpha}{{\Gamma[1+\alpha]}}\big)}}+e^{{\big({\frac{2kx {\Gamma[1+\alpha]}-2\lambda t^\alpha}{{\Gamma[1+\alpha]}}\big)}}}}{1+2 e^{\big({\frac{kx {\Gamma[1+\alpha]}-\lambda t^\alpha}{{\Gamma[1+\alpha]}}\big)}}+e^{\big({\frac{2kx {\Gamma[1+\alpha]}-2\lambda t^\alpha}{{\Gamma[1+\alpha]}}\big)}}},
\end{equation}
then the exact solution is 
\begin{equation} \label{expratcase1}
    u(x,t)=\displaystyle \left [\frac{2 e^{\big({\frac{kx {\Gamma[1+\alpha]}-\lambda t^\alpha}{{\Gamma[1+\alpha]}}\big)}}+e^{{\big({\frac{2kx {\Gamma[1+\alpha]}-2\lambda t^\alpha}{{\Gamma[1+\alpha]}}\big)}}}}{1+2 e^{\big({\frac{kx {\Gamma[1+\alpha]}-\lambda t^\alpha}{{\Gamma[1+\alpha]}}\big)}}+e^{\big({\frac{2kx {\Gamma[1+\alpha]}-2\lambda t^\alpha}{{\Gamma[1+\alpha]}}\big)}}}\right]^ \frac{1}{\delta}.
\end{equation}
Case: 2\\
\[a_0=1,\; a_1=-1, \;a_2=-1-a_1,\;  \gamma=-\frac{2 \beta^2 \delta}{(1+\delta)^2}, \;\lambda=k^2+k\beta+\gamma\delta,\; k=-\frac{\beta \delta}{1+\delta}, \]
Eq.(\ref{nonlinearode}) transforms into
\begin{equation}
    \delta v v^{\prime \prime}+(1+\delta)v^2v^{\prime}+(2\delta-1)v v^{\prime}-(\delta-1){v^\prime}^2+2\delta v^2(v-1)=0,
\end{equation}
we get
\begin{equation}
    v(x,t)=\displaystyle \frac{e^{\big({\frac{kx {\Gamma[1+\alpha]}-\lambda t^\alpha}{{2\Gamma[1+\alpha]}}\big)}}}{2 cosh \Big( {\frac{kx {\Gamma[1+\alpha]}-\lambda t^\alpha}{{2\Gamma[1+\alpha]}} \Big) }},
\end{equation}
then the exact solution is 
\begin{equation} \label{expratcase2}
    u(x,t)=\displaystyle \left[\frac{e^{\big({\frac{kx {\Gamma[1+\alpha]}-\lambda t^\alpha}{{2\Gamma[1+\alpha]}}\big)}}}{2 cosh\Big({\frac{kx {\Gamma[1+\alpha]}-\lambda t^\alpha}{{2\Gamma[1+\alpha]}}\Big)}}\right]^ \frac{1}{\delta}.
\end{equation}
Case: 3\\
\[a_0=1,\; a_1=-2, \;a_2=-1-a_1,\; \beta=\frac{-4k^2-2k^2\delta+\gamma\delta^2}{4k\delta},\; \gamma=-\frac{2 k^2(2+\delta)}{\delta^2},\] 
\[\;\lambda=k^2+k\beta+\gamma\delta,\; k=-\frac{\beta \delta}{1+\delta}, \]
Eq.(\ref{nonlinearode}) becomes
\begin{equation}
    \delta v v^{\prime \prime}-(4+\delta)v v^{\prime}-(\delta-1){v^\prime}^2 -2(2+\delta) (1-v) v^2=0,
\end{equation}
we obtain
\begin{equation}
    v(x,t)=\frac{e^{\big({\frac{2kx {\Gamma[1+\alpha]}-2\lambda t^\alpha}{{\Gamma[1+\alpha]}}\big)}}}{1+2 e^{\big({\frac{kx {\Gamma[1+\alpha]}-\lambda t^\alpha}{{\Gamma[1+\alpha]}}\big)}}+e^{\big({\frac{2kx {\Gamma[1+\alpha]}-2\lambda t^\alpha}{{\Gamma[1+\alpha]}}\big)}}},
\end{equation}
then the exact solution is 
\begin{equation} \label{expratcase3}
    u(x,t)=\displaystyle \left[\frac{e^{\big({\frac{2kx {\Gamma[1+\alpha]}-2\lambda t^\alpha}{{\Gamma[1+\alpha]}}\big)}}}{1+2 e^{\big({\frac{kx {\Gamma[1+\alpha]}-\lambda t^\alpha}{{\Gamma[1+\alpha]}}\big)}}+e^{\big({\frac{2kx {\Gamma[1+\alpha]}-2\lambda t^\alpha}{{\Gamma[1+\alpha]}}\big)}}}\right]^ \frac{1}{\delta}.
\end{equation}
Case: 4\\
\[a_0=0,\; a_1=-a_2, \;a_2=\frac{-2k(2+\delta)}{k-\beta \delta},\; \beta=0,\; \gamma=-\frac{ k^2}{\delta^2}, 
\;\lambda=\frac{k^2+\gamma\delta^2}{\delta}, \]
Eq.(\ref{nonlinearode}) becomes
\begin{equation}
    \delta v v^{\prime \prime}-(\delta-1){v^\prime}^2+v^3- v^2=0,
\end{equation}
we obtain
\begin{equation}
    v(x,t)=\frac{4e^{\big({\frac{kx {\Gamma[1+\alpha]}-\lambda t^\alpha}{{\Gamma[1+\alpha]}}\big)}}+4\delta e^{\big({\frac{kx {\Gamma[1+\alpha]}-\lambda t^\alpha}{{\Gamma[1+\alpha]}}\big)}}}{1+2 e^{\big({\frac{kx {\Gamma[1+\alpha]}-\lambda t^\alpha}{{\Gamma[1+\alpha]}}\big)}}+e^{\big({\frac{2kx {\Gamma[1+\alpha]}-2\lambda t^\alpha}{{\Gamma[1+\alpha]}}\big)}}},
\end{equation}
then the exact solution is 
\begin{equation} \label{expratcase4}
    u(x,t)=\displaystyle \left[\frac{4e^{\big({\frac{kx {\Gamma[1+\alpha]}-\lambda t^\alpha}{{\Gamma[1+\alpha]}}\big)}}+4\delta e^{\big({\frac{kx {\Gamma[1+\alpha]}-\lambda t^\alpha}{{\Gamma[1+\alpha]}}\big)}}}{1+2 e^{\big({\frac{kx {\Gamma[1+\alpha]}-\lambda t^\alpha}{{\Gamma[1+\alpha]}}\big)}}+e^{\big({\frac{2kx {\Gamma[1+\alpha]}-2\lambda t^\alpha}{{\Gamma[1+\alpha]}}\big)}}}\right]^ \frac{1}{\delta}.
\end{equation}
Case: 5\\
\[a_0=0,\; a_1=1-a_2, \;a_2=0,\;  \gamma=\frac{ \beta^2}{1+\delta}, 
\;\lambda=\frac{k^2+\gamma\delta^2}{\delta},\;k=\frac{\beta \delta}{1+\delta}, \]
Eq.(\ref{nonlinearode}) becomes
\begin{equation}
    -\delta v v^{\prime \prime}-(2+\delta)v{v^\prime}+(\delta-1){v^\prime}^2+(1+\delta) v^{\prime}v^2-(1+\delta)(1+v)v^2=0,
\end{equation}
we obtain
\begin{equation}
     v(x,t)=\frac{e^{-\big({\frac{kx {\Gamma[1+\alpha]}-\lambda t^\alpha}{{2\Gamma[1+\alpha]}}\big)}}}{2 cosh \;e^{\Big({\frac{kx {\Gamma[1+\alpha]}-\lambda t^\alpha}{{2\Gamma[1+\alpha]}}\Big)}}},
\end{equation}
then the exact solution is 
\begin{equation} \label{expratcase5}
    u(x,t)=\displaystyle \left[\frac{e^{-\big({\frac{kx {\Gamma[1+\alpha]}-\lambda t^\alpha}{{2\Gamma[1+\alpha]}}\big)}}}{2 cosh \;e^{\Big({\frac{kx {\Gamma[1+\alpha]}-\lambda t^\alpha}{{2\Gamma[1+\alpha]}}\Big)}}}\right]^ \frac{1}{\delta}.
\end{equation}
Case: 6\\
\[a_0=0,\; a_1=1-a_2, \;a_2=-1,\; \beta=-\frac{k(\delta-1)}{\delta},\; \gamma=-\frac{2k(k+2k\delta-\beta \delta)}{\delta^2}, 
\;\]
\[\lambda=\frac{k^2+\gamma\delta^2}{\delta},\;\delta=1, \]
Eq.(\ref{nonlinearode}) becomes
\begin{equation}
        v^{\prime \prime}-5v^{\prime}+6v^2-6v=0,
\end{equation}
we obtain
\begin{equation}
    v(x,t)=\frac{1+2 e^{\big({\frac{kx {\Gamma[1+\alpha]}-\lambda t^\alpha}{{\Gamma[1+\alpha]}}\big)}}}{1+2 e^{\big({\frac{kx {\Gamma[1+\alpha]}-\lambda t^\alpha}{{\Gamma[1+\alpha]}}\big)}}+e^{\big({\frac{2kx {\Gamma[1+\alpha]}-2\lambda t^\alpha}{{\Gamma[1+\alpha]}}\big)}}},
\end{equation}
then the exact solution is 
\begin{equation} \label{expratcase6}
    u(x,t)=\displaystyle \left[\frac{1+2 e^{\big({\frac{kx {\Gamma[1+\alpha]}-\lambda t^\alpha}{{\Gamma[1+\alpha]}}\big)}}}{1+2 e^{\big({\frac{kx {\Gamma[1+\alpha]}-\lambda t^\alpha}{{\Gamma[1+\alpha]}}\big)}}+e^{\big({\frac{2kx {\Gamma[1+\alpha]}-2\lambda t^\alpha}{{\Gamma[1+\alpha]}}\big)}}}\right]^ \frac{1}{\delta}.
\end{equation}
Case: 7\\
\[a_0=0,\; a_1=0, \;a_2=1,\; \beta=\frac{4k^2+k^2\delta-\delta \lambda}{k\delta},\; \lambda=\frac{k^2(4+\delta)}{\delta},\]
Eq.(\ref{nonlinearode}) becomes
\begin{equation}
    \delta v v^{\prime \prime}+(4+\delta)v v^{\prime}-(\delta-1){v^\prime}^2 -2(2+\delta) (v-1) v^2=0,
\end{equation}
we obtain
\begin{equation}
    v(x,t)=\frac{1}{1+2 e^{\big({\frac{kx {\Gamma[1+\alpha]}-\lambda t^\alpha}{{\Gamma[1+\alpha]}}\big)}}+e^{\big({\frac{2kx {\Gamma[1+\alpha]}-2\lambda t^\alpha}{{\Gamma[1+\alpha]}}\big)}}},
\end{equation}
then the exact solution is 
\begin{equation} \label{expratcase7}
    u(x,t)=\displaystyle \left[\frac{1}{1+2 e^{\big({\frac{kx {\Gamma[1+\alpha]}-\lambda t^\alpha}{{\Gamma[1+\alpha]}}\big)}}+e^{\big({\frac{2kx {\Gamma[1+\alpha]}-2\lambda t^\alpha}{{\Gamma[1+\alpha]}}\big)}}}\right]^ \frac{1}{\delta}.
\end{equation}

\section{Results and Discussion}
The exact solutions of the TF-GBF have been generated by the Exp function method. The 2D and 3D plots show the exact solutions through Figures (1)–(9) with $\alpha=0.25,0.5,0.65,$ and $0.78$ for different values of space variable $x$ and time $t$. These figures show when $x$ and $t$ increase, the solution $u$  decreases for the corresponding exact solutions. Fig (9) shows, when  $x$ and $t$ increase, the solution $u$ decreases for the exact solution of Eq. (\ref{expfuncase9}). 

The ERFM was used to obtain the exact solution of the TF-GBF.  The 2D and 3D plots show the exact solutions via Figures (10)–(16) with $\alpha=0.25,\;0.35,\;0.41,\;0.45,\;0.5,\;0.68,$ and $0.7$, for various space variable $x$ and time $t$. Fig (10) and (12) display the 2D and 3D plots of Eq.(\ref{expratcase1}) and Eq.(\ref{expratcase3}) respectively which shows the solution $u$ increases, when $x$ and $t$ increase. Fig (11), (14), (15) and (16) display the 2D and 3D plots of  Eq.(\ref{expratcase2}), Eq.(\ref{expratcase5}), Eq.(\ref{expratcase6}) and Eq.(\ref{expratcase7}) respectively as $x$ and $t$ increases, the solution $u$ decreases.  Fig.(13) shows the 2D and 3D plot of Eq. (\ref{expratcase4}) when  $x$ and $t$ increase or decrease, the solution  $u$ also increases or decreases.

\begin{figure}
\includegraphics[width=0.475\textwidth]{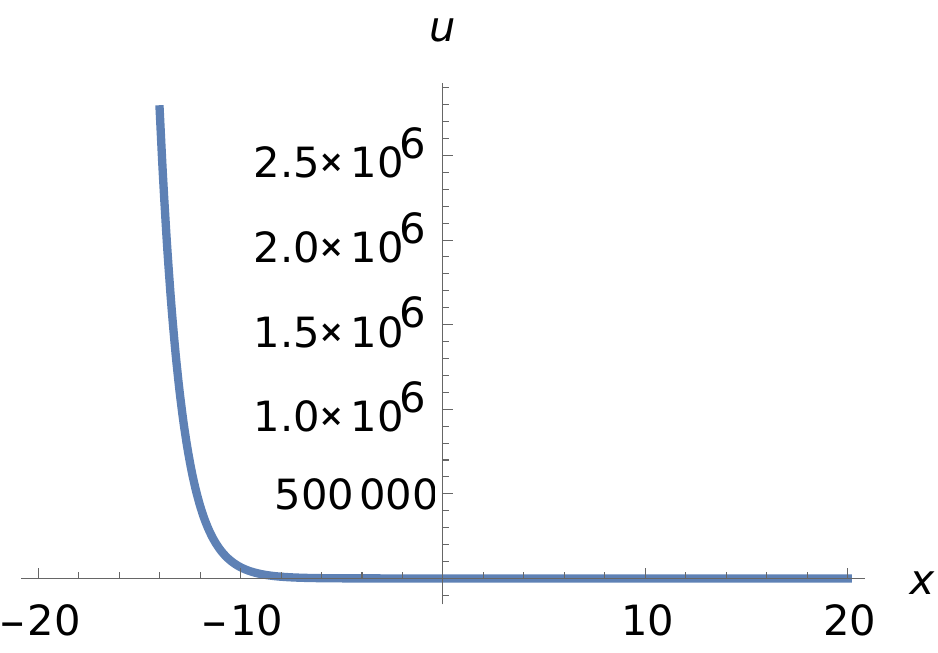}
\hspace{\fill}
\includegraphics[width=0.475\textwidth]{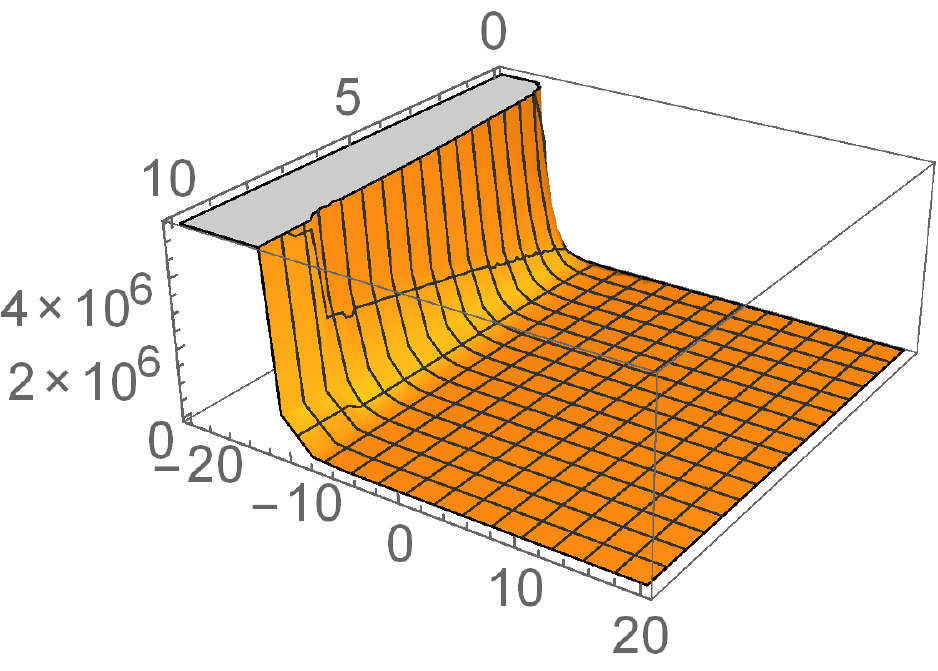}
\caption{2D and 3D shape of  Eq.(\ref{expfuncase1}) }\label{fig:1}
\end{figure}

\begin{figure}
\includegraphics[width=0.475\textwidth]{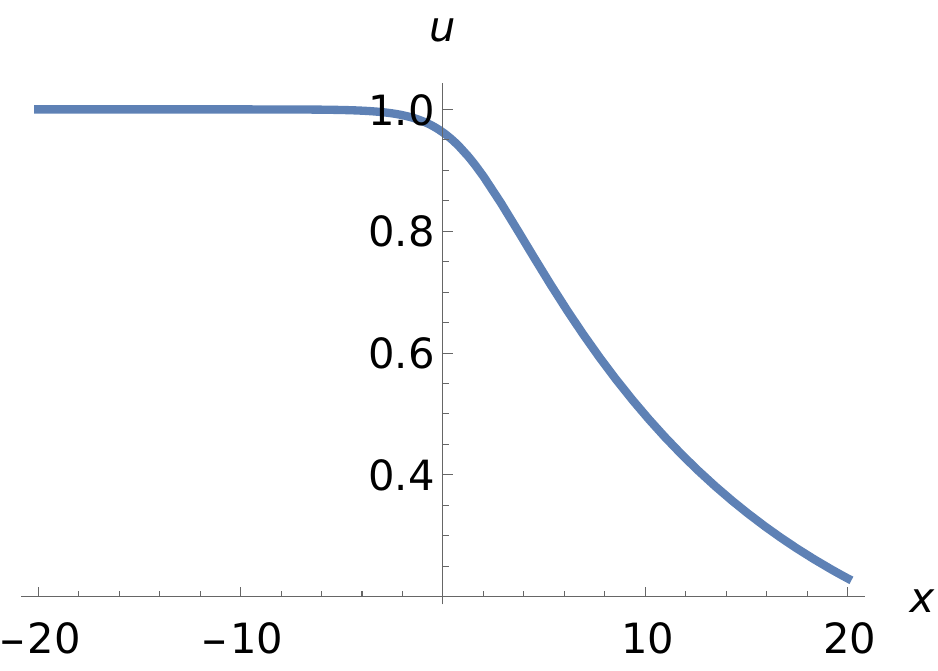}
\hspace{\fill}
\includegraphics[width=0.475\textwidth]{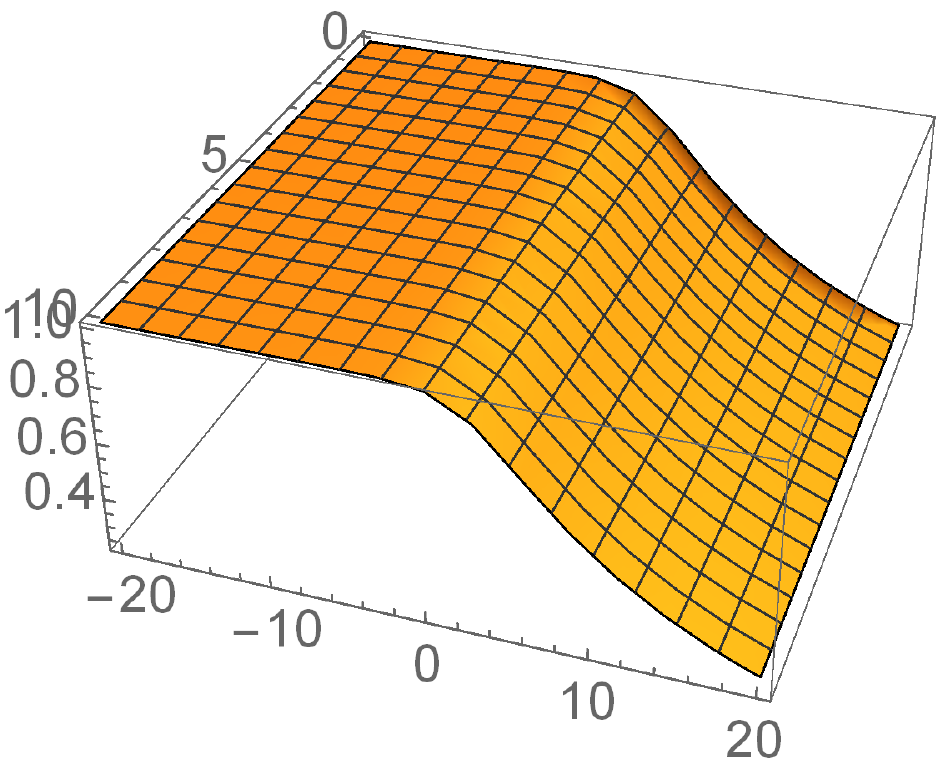}
\caption{2D and 3D shape of  Eq.(\ref{expfuncase2}) }\label{fig:2}
\end{figure}

\begin{figure}
\includegraphics[width=0.475\textwidth]{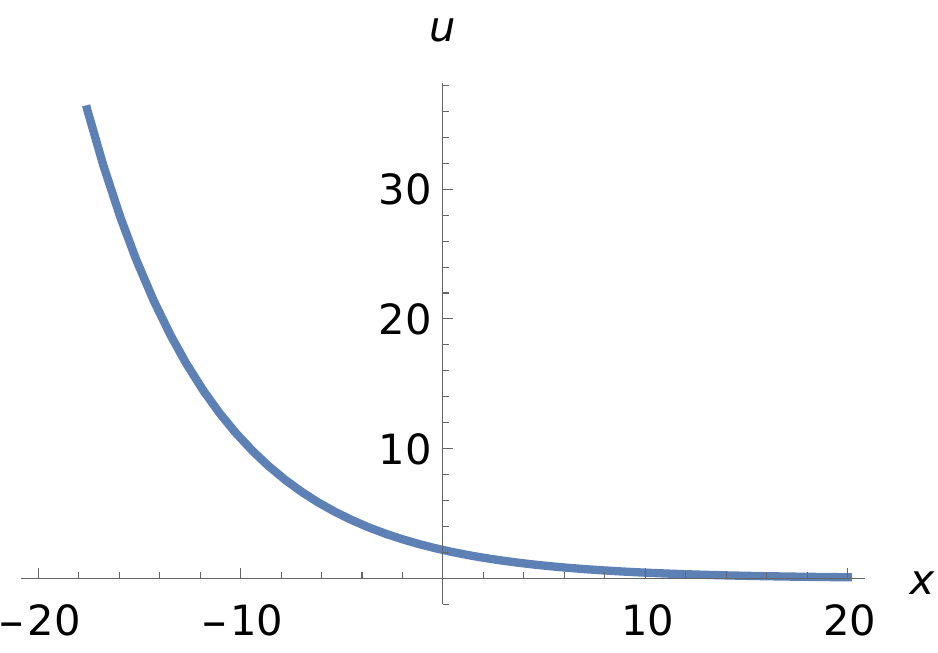}
\hspace{\fill}
\includegraphics[width=0.475\textwidth]{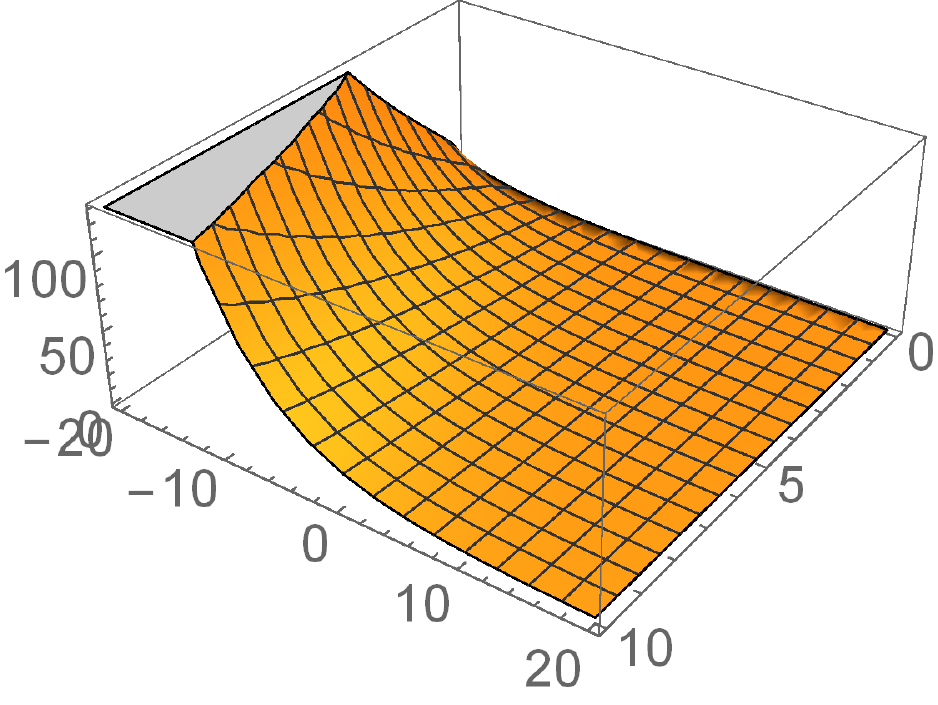}
\caption{2D and 3D shape of  Eq.(\ref{expfuncase3}) }\label{fig:3}
\end{figure}

\begin{figure}
\includegraphics[width=0.475\textwidth]{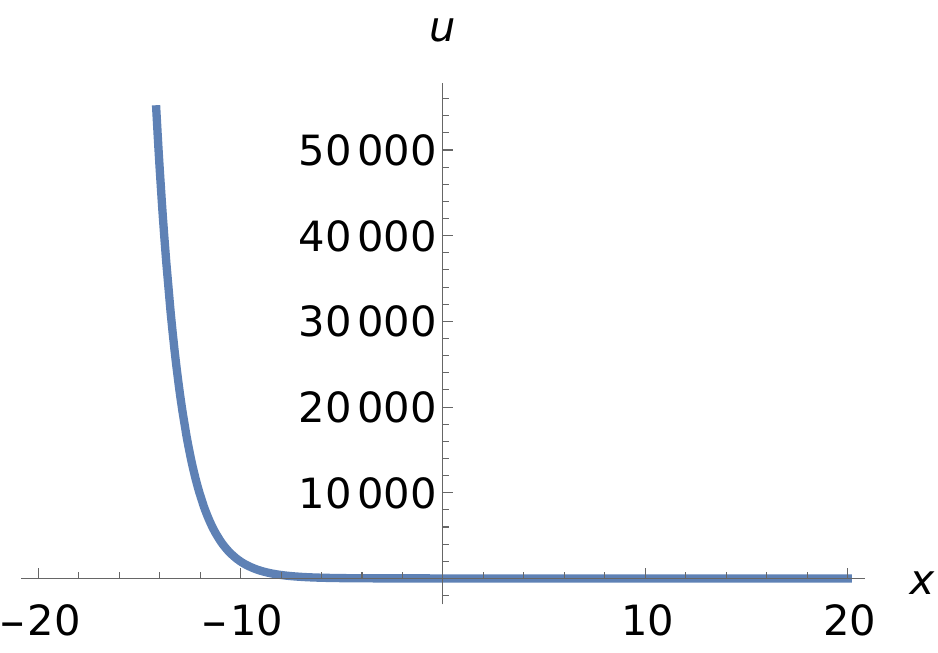}
\hspace{\fill}
\includegraphics[width=0.475\textwidth]{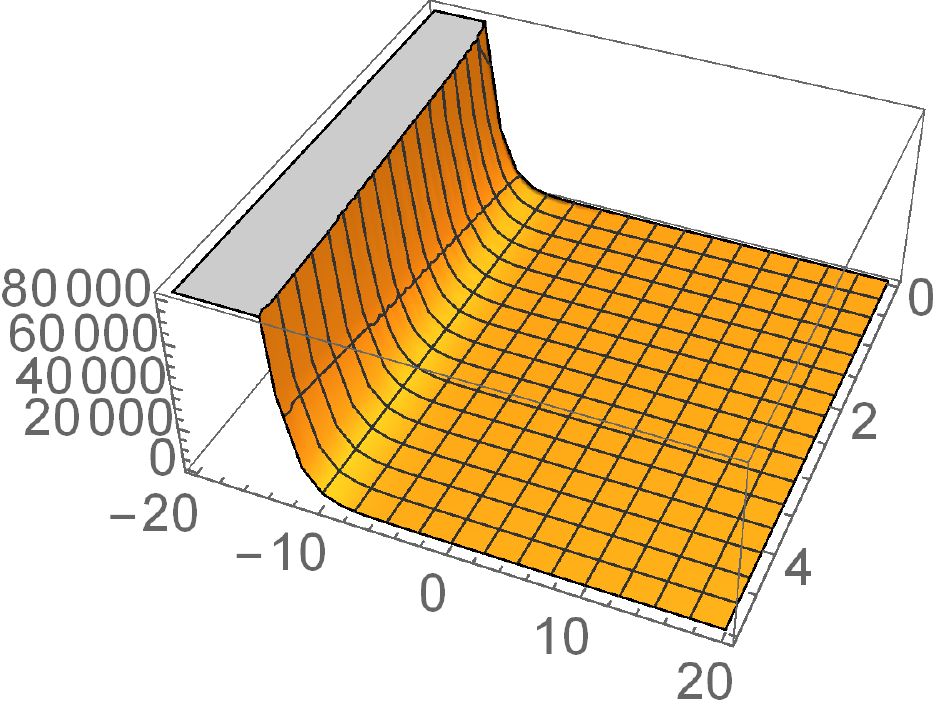}
\caption{2D and 3D shape of  Eq.(\ref{expfuncase4}) }\label{fig:4}
\end{figure}

\begin{figure}
\includegraphics[width=0.475\textwidth]{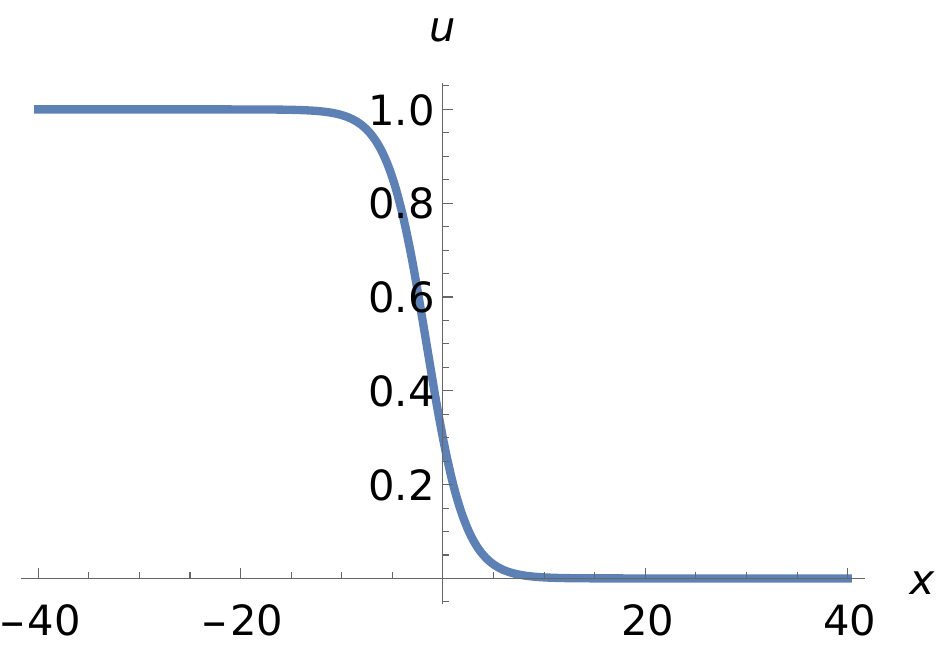}
\hspace{\fill}
\includegraphics[width=0.475\textwidth]{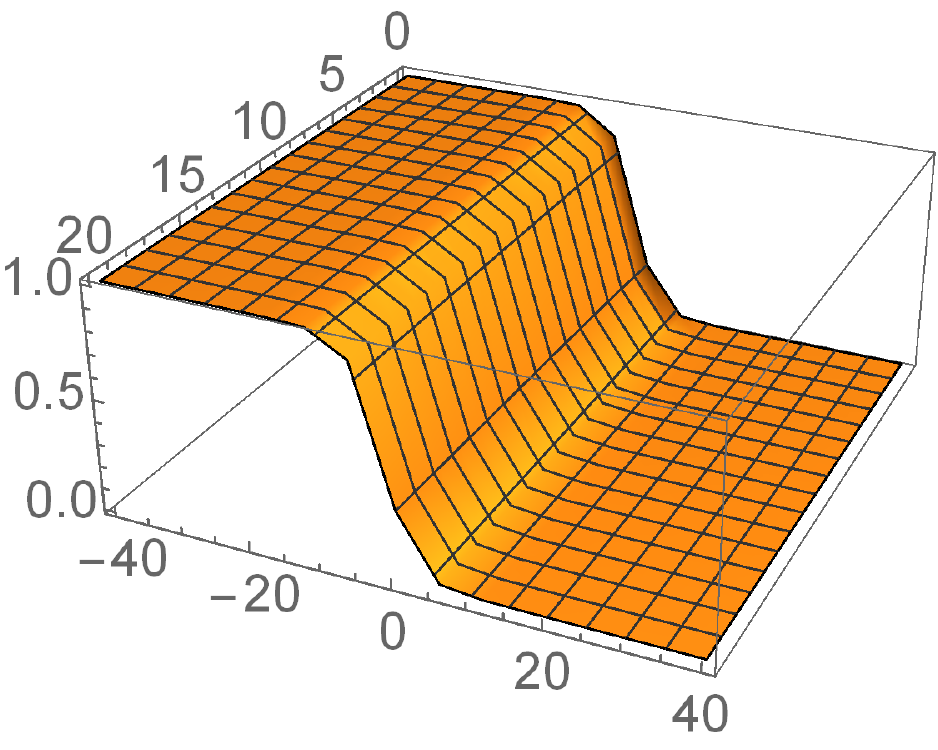}
\caption{2D and 3D shape of  Eq.(\ref{expfuncase5}) }\label{fig:5}
\end{figure}

\begin{figure}
\includegraphics[width=0.475\textwidth]{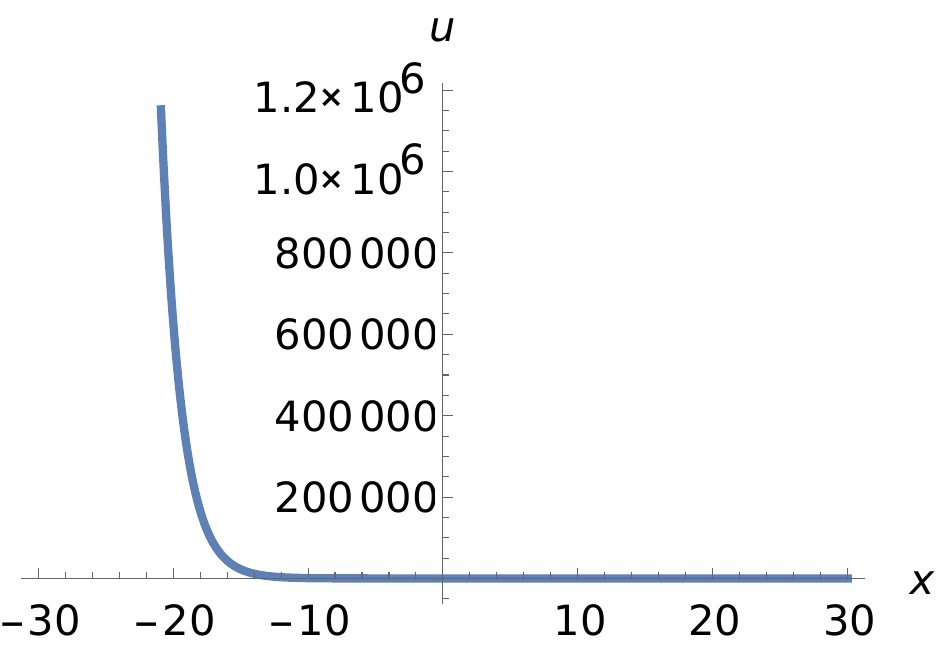}
\hspace{\fill}
\includegraphics[width=0.475\textwidth]{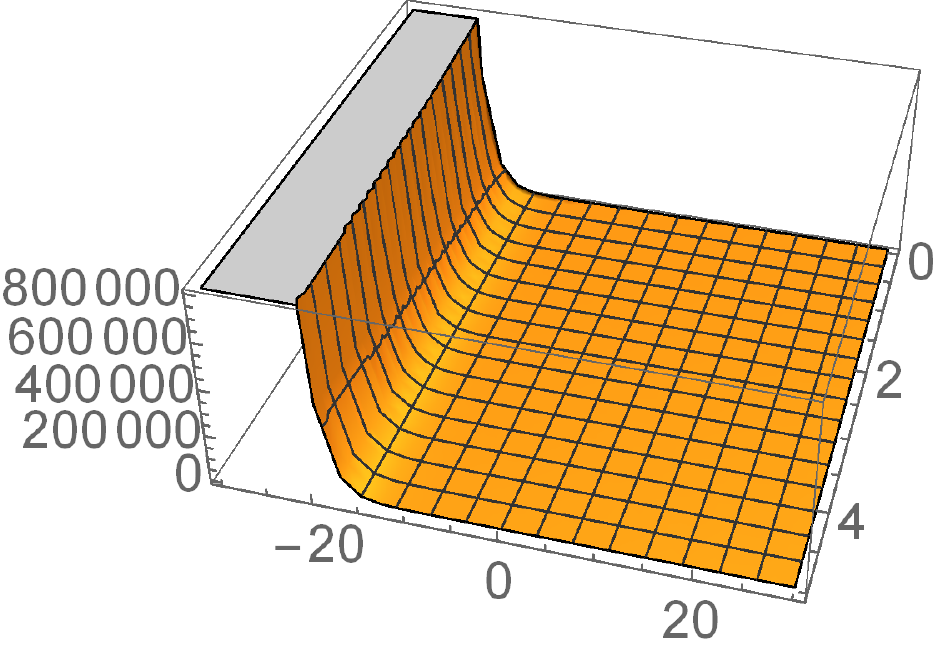}
\caption{2D and 3D shape of  Eq.(\ref{expfuncase6}) }\label{fig:6}
\end{figure}

\begin{figure}
\includegraphics[width=0.475\textwidth]{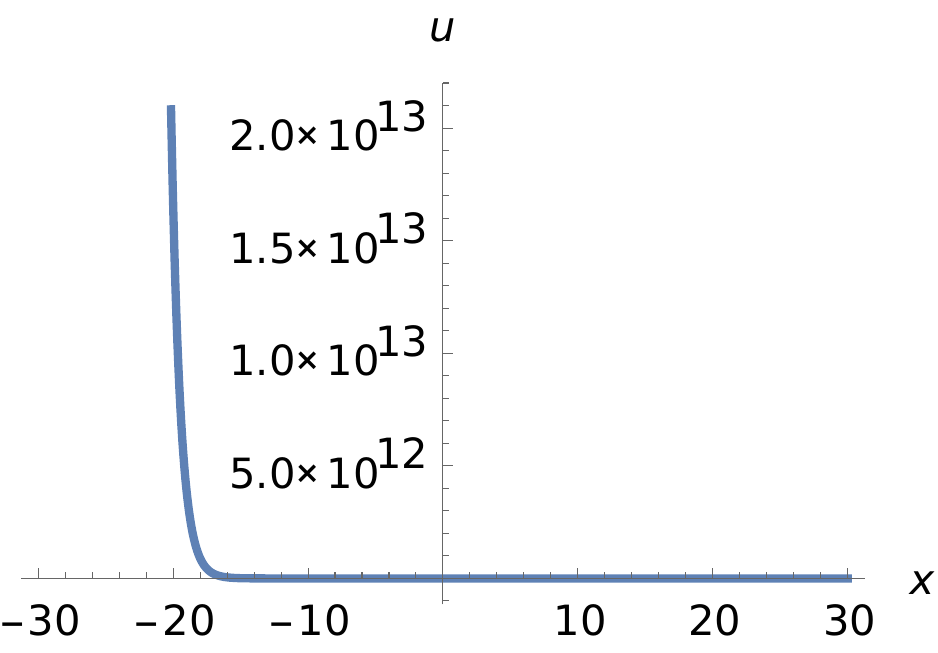}
\hspace{\fill}
\includegraphics[width=0.475\textwidth]{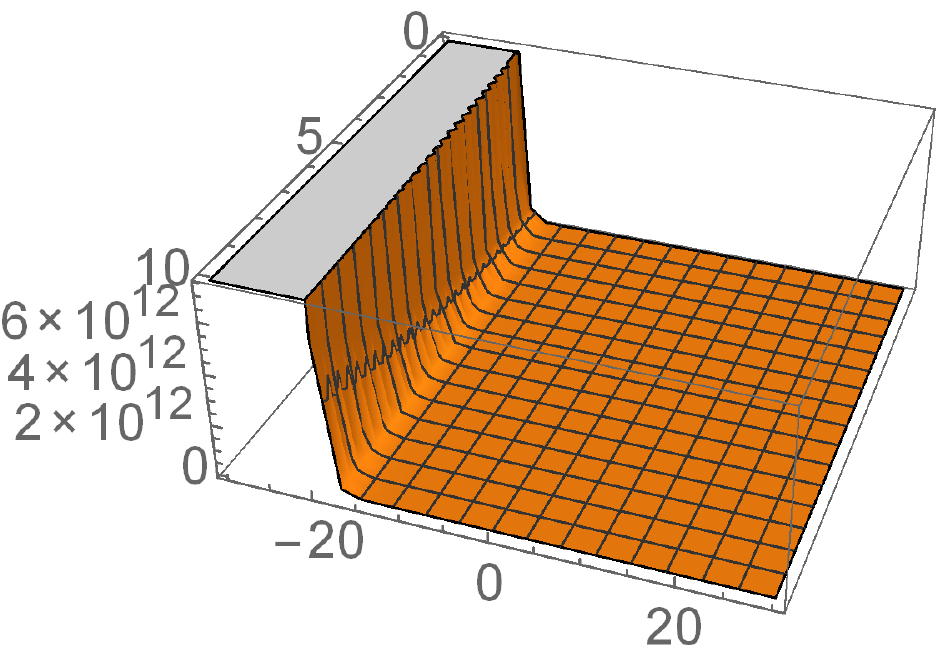}
\caption{2D and 3D shape of  Eq.(\ref{expfuncase7}) }\label{fig:7}
\end{figure}

\begin{figure}
\includegraphics[width=0.475\textwidth]{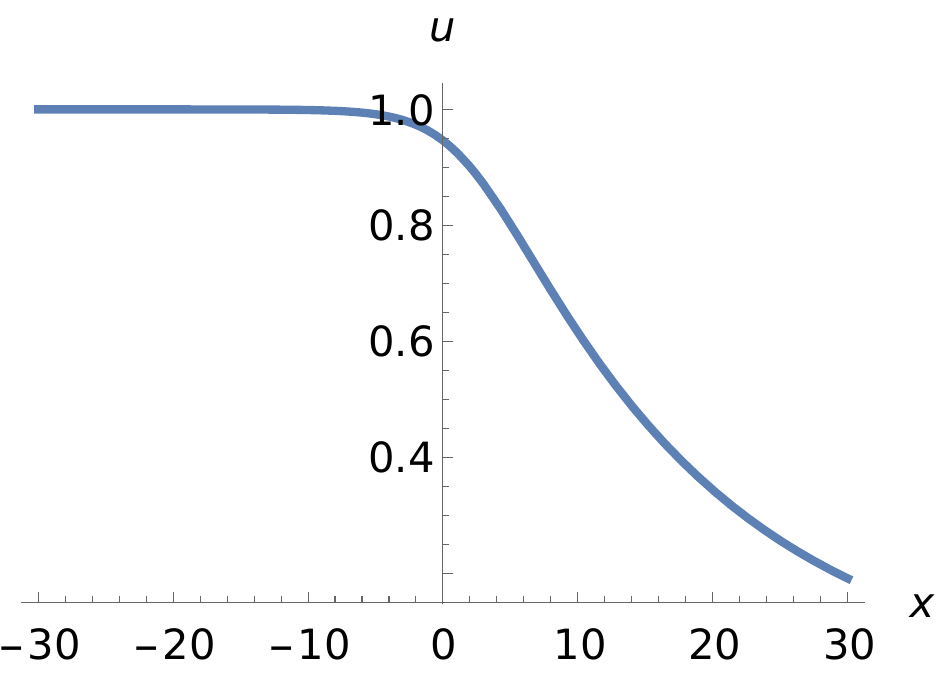}
\hspace{\fill}
\includegraphics[width=0.475\textwidth]{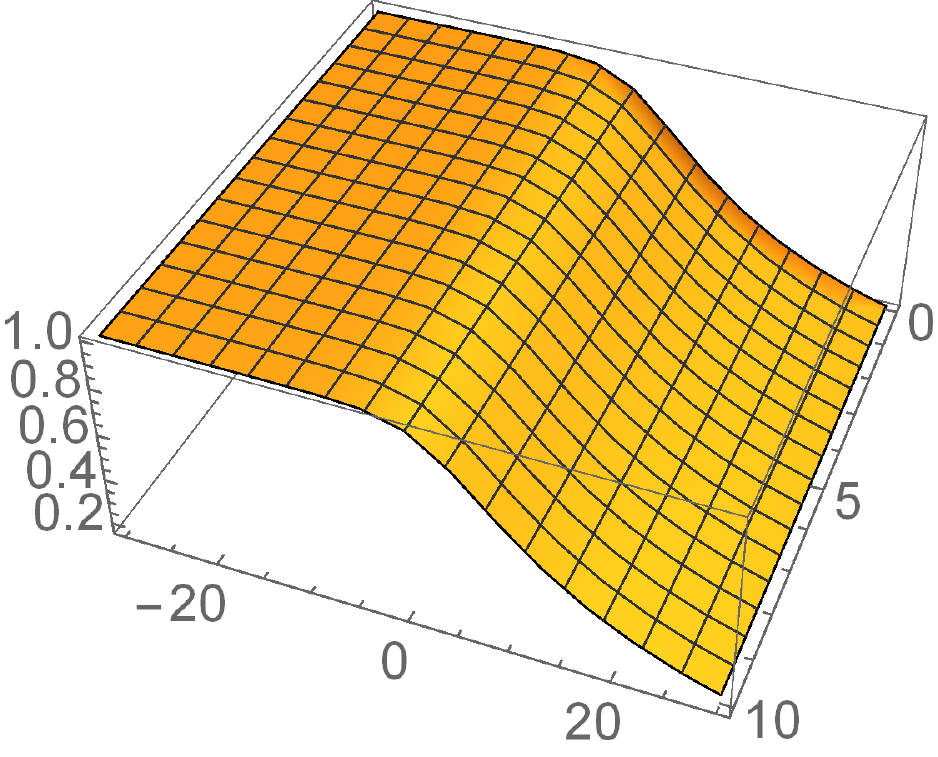}
\caption{2D and 3D shape of  Eq.(\ref{expfuncase8}) }\label{fig:8}
\end{figure}

\begin{figure}
\includegraphics[width=0.475\textwidth]{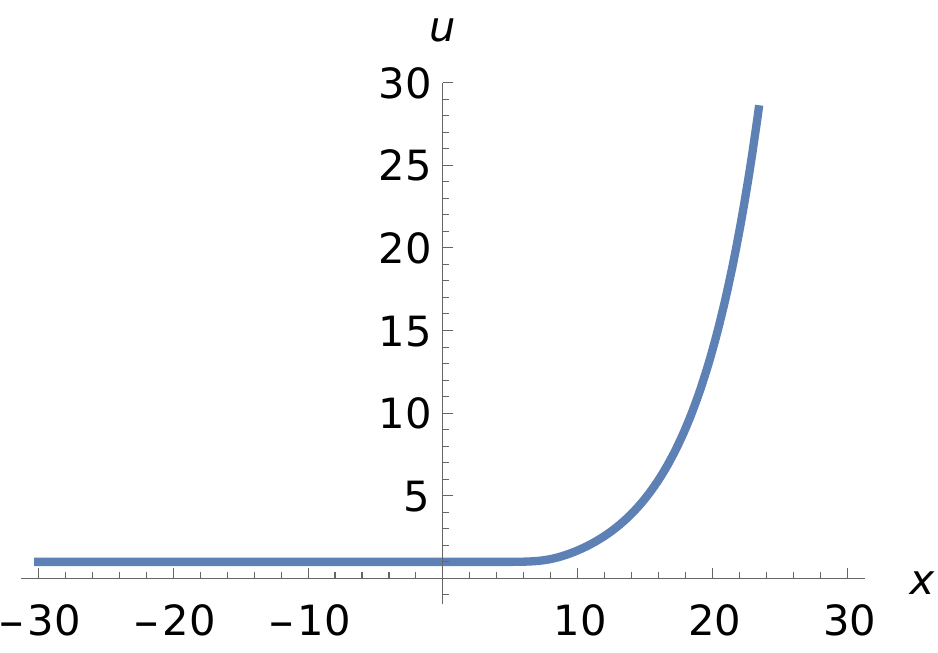}
\hspace{\fill}
\includegraphics[width=0.475\textwidth]{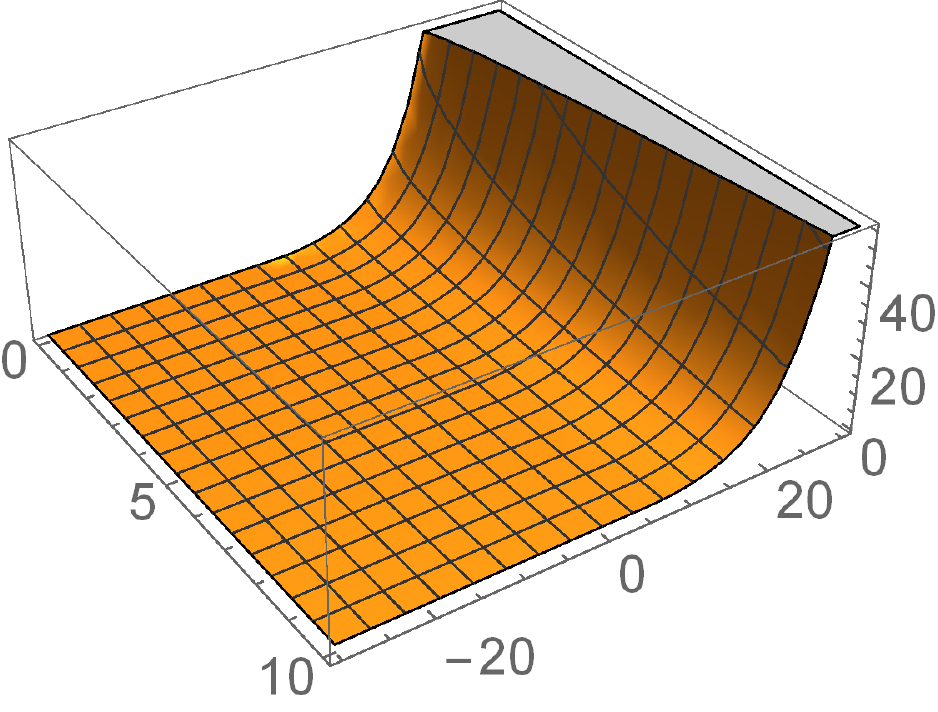}
\caption{2D and 3D shape of  Eq.(\ref{expfuncase9}) }\label{fig:9}
\end{figure}

\begin{figure}
\includegraphics[width=0.475\textwidth]{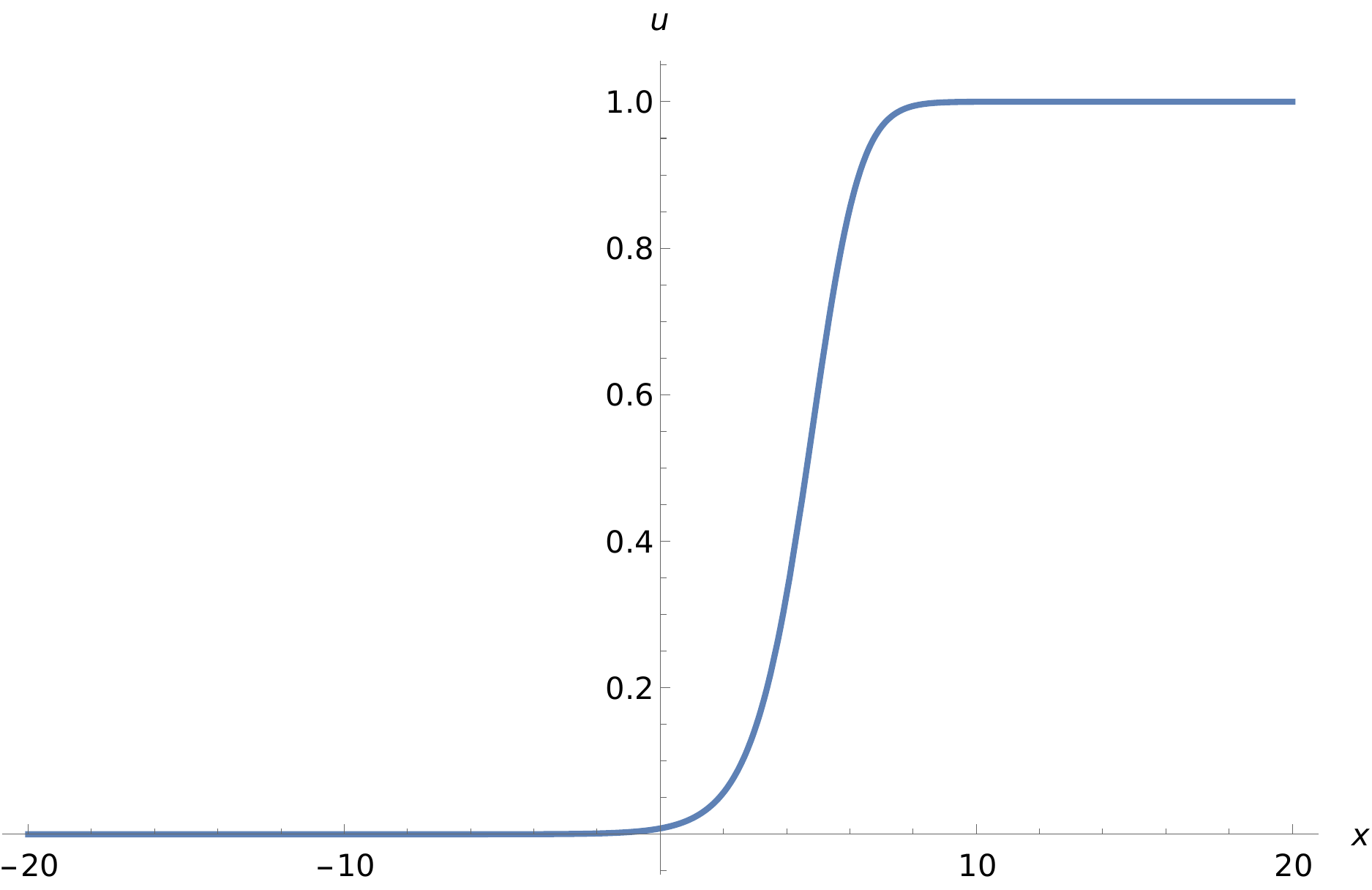}
\hspace{\fill}
\includegraphics[width=0.475\textwidth]{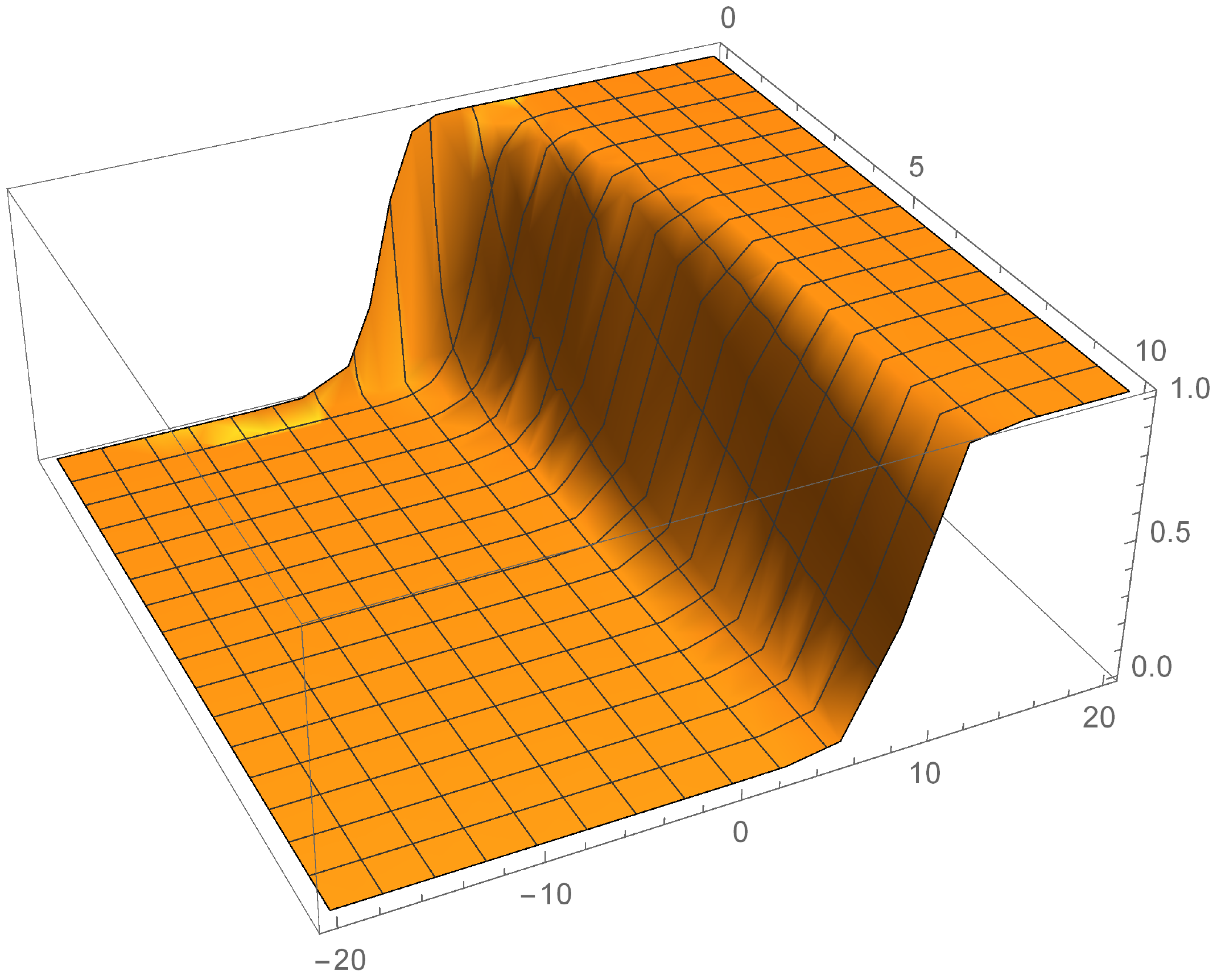}
\caption{2D and 3D shape of  Eq.(\ref{expratcase1}) }\label{fig:10}
\end{figure}

\begin{figure}
\includegraphics[width=0.475\textwidth]{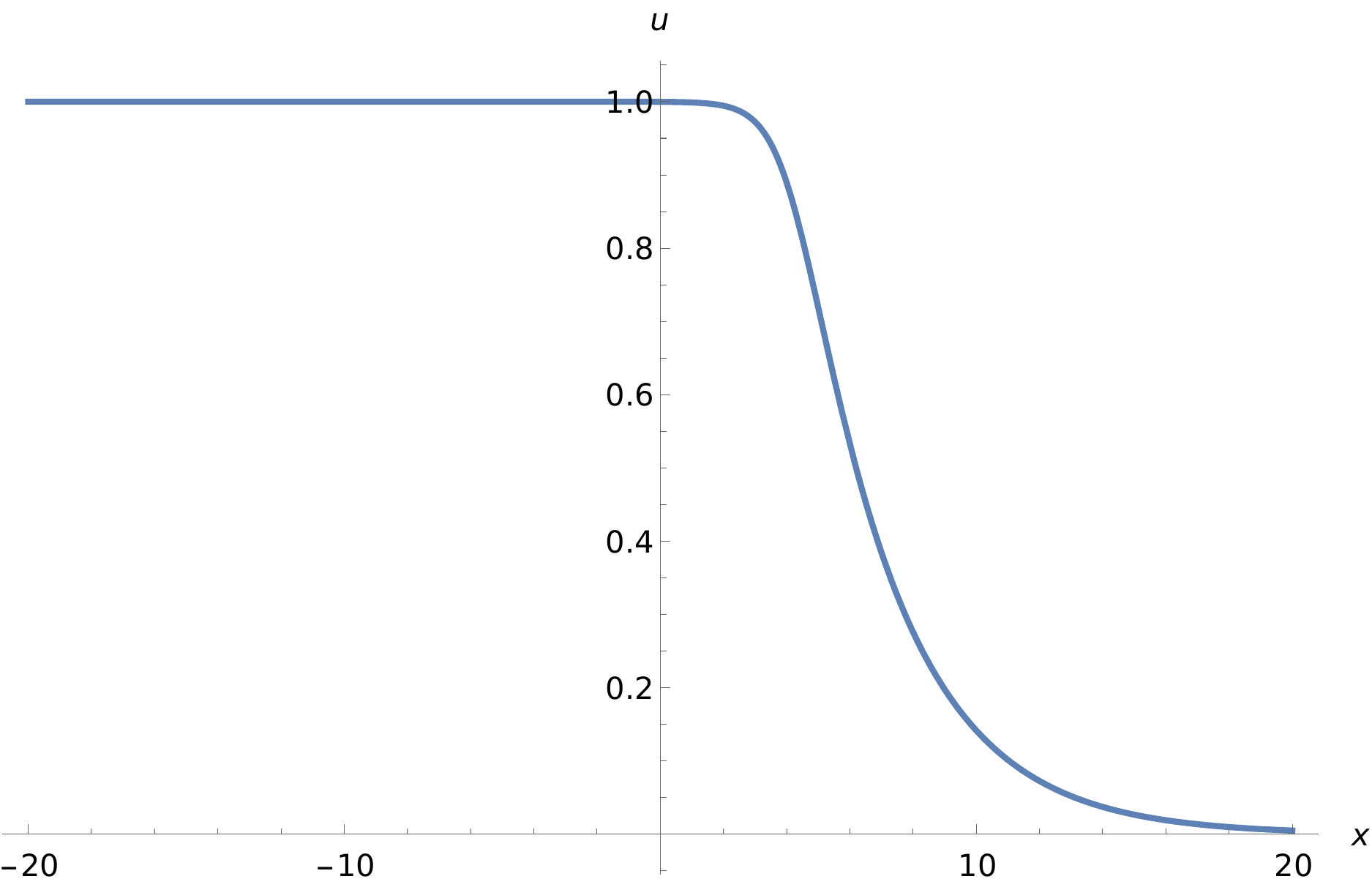}
\hspace{\fill}
\includegraphics[width=0.475\textwidth]{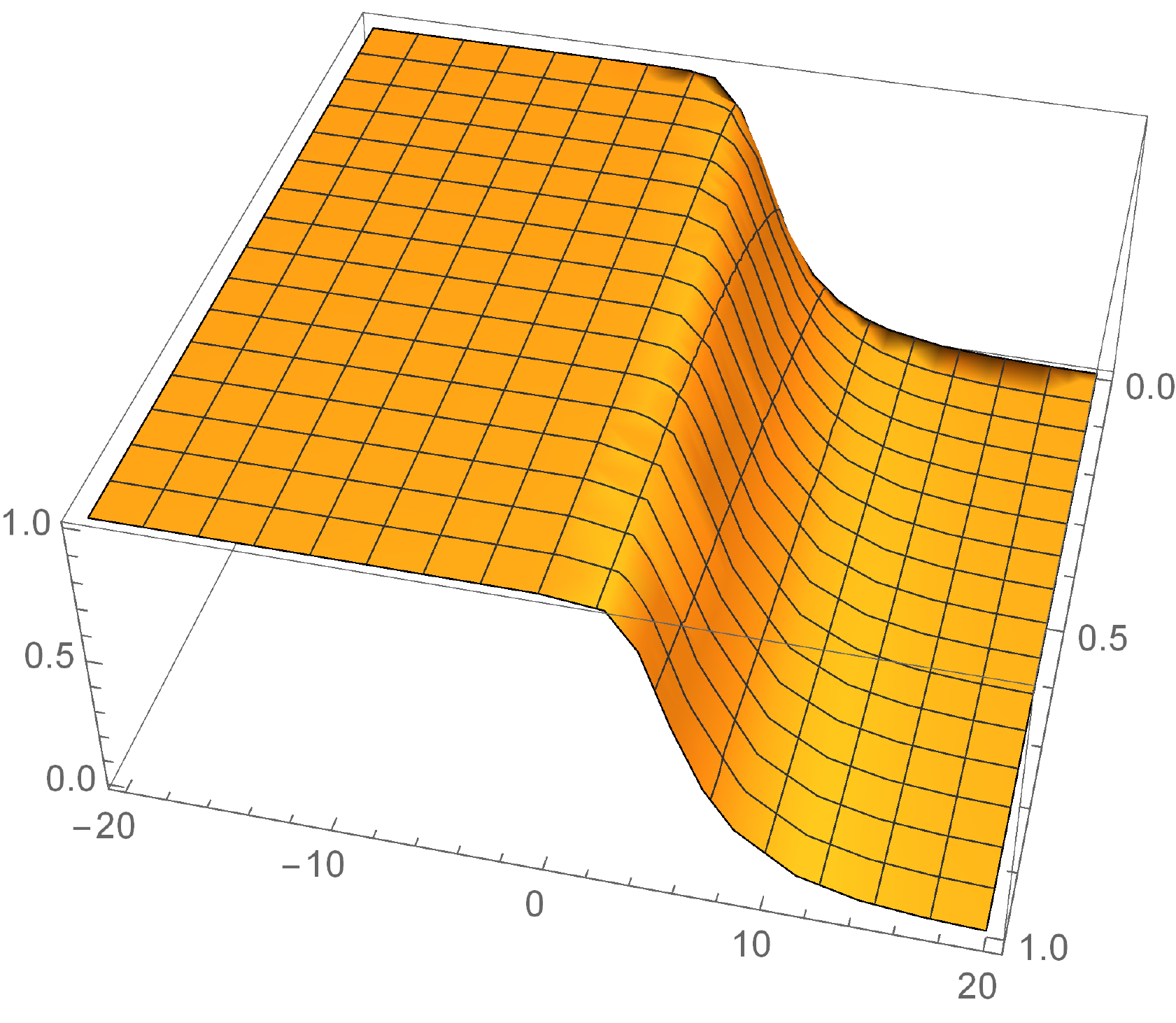}
\caption{2D and 3D shape of  Eq.(\ref{expratcase2}) }\label{fig:11}
\end{figure}

\begin{figure}
\includegraphics[width=0.47\textwidth]{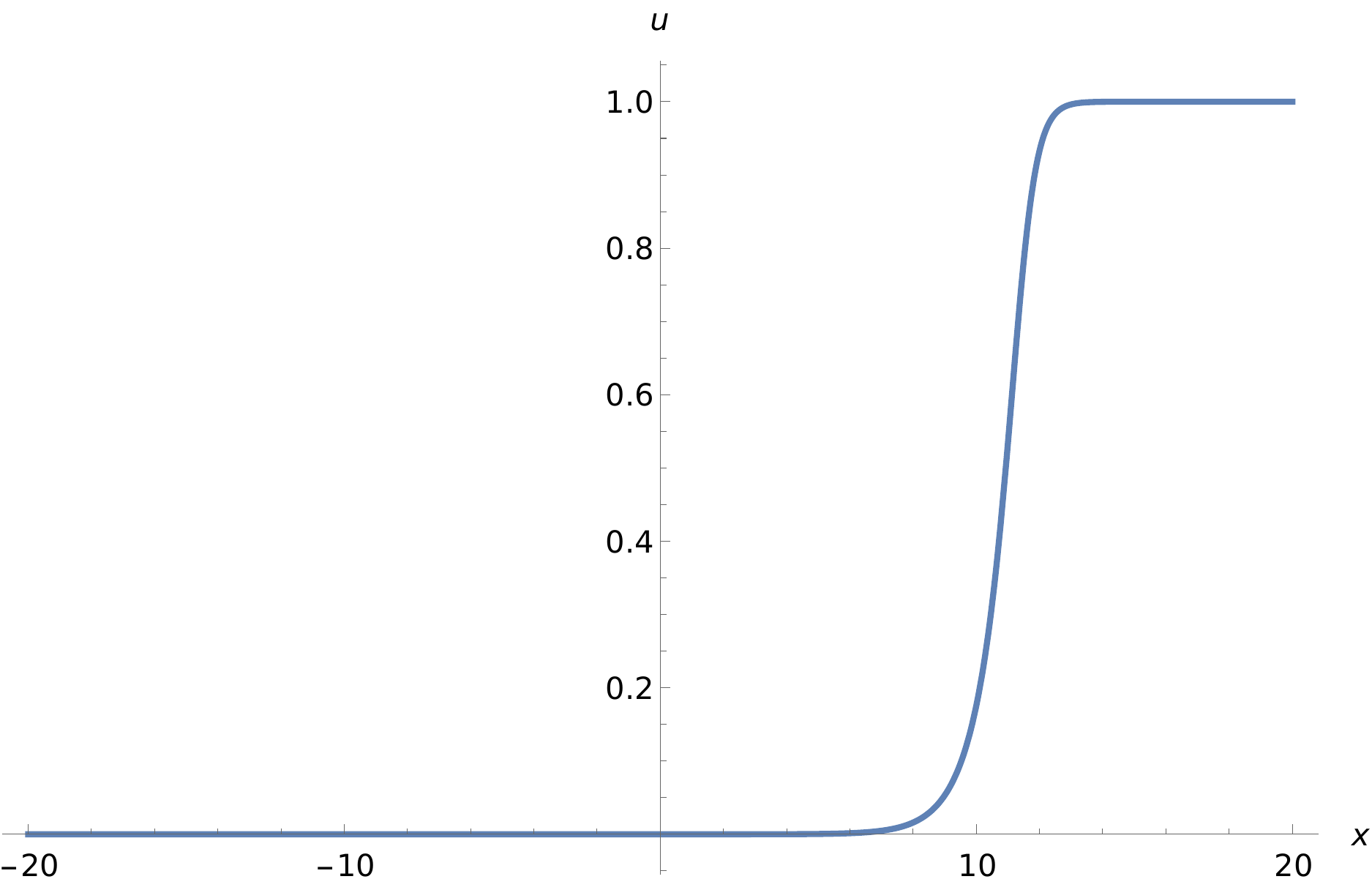}
\hspace{\fill}
\includegraphics[width=0.47\textwidth]{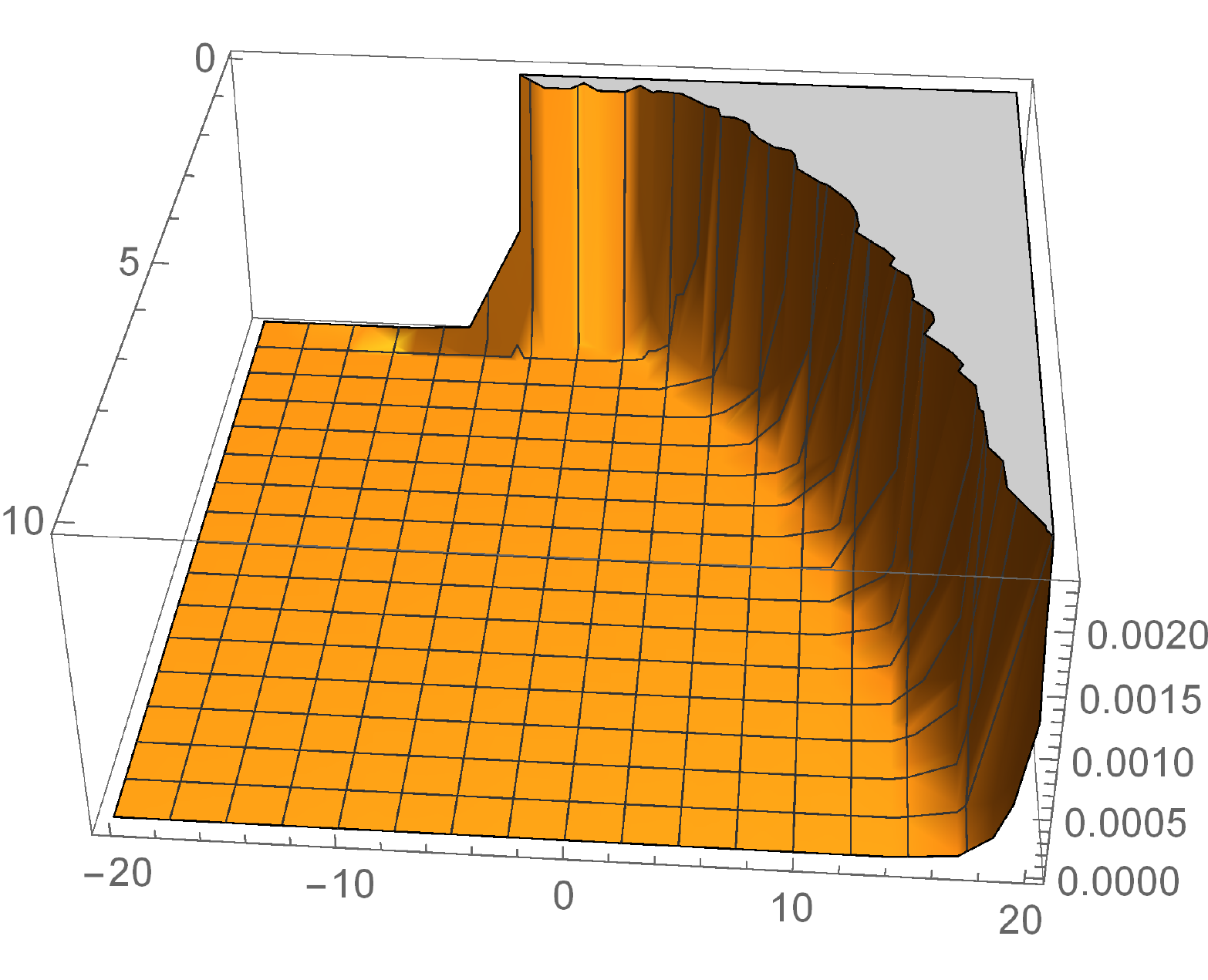}
\caption{2D and 3D shape of  Eq.(\ref{expratcase3}) }\label{fig:12}
\end{figure}

\begin{figure}
\includegraphics[width=0.475\textwidth]{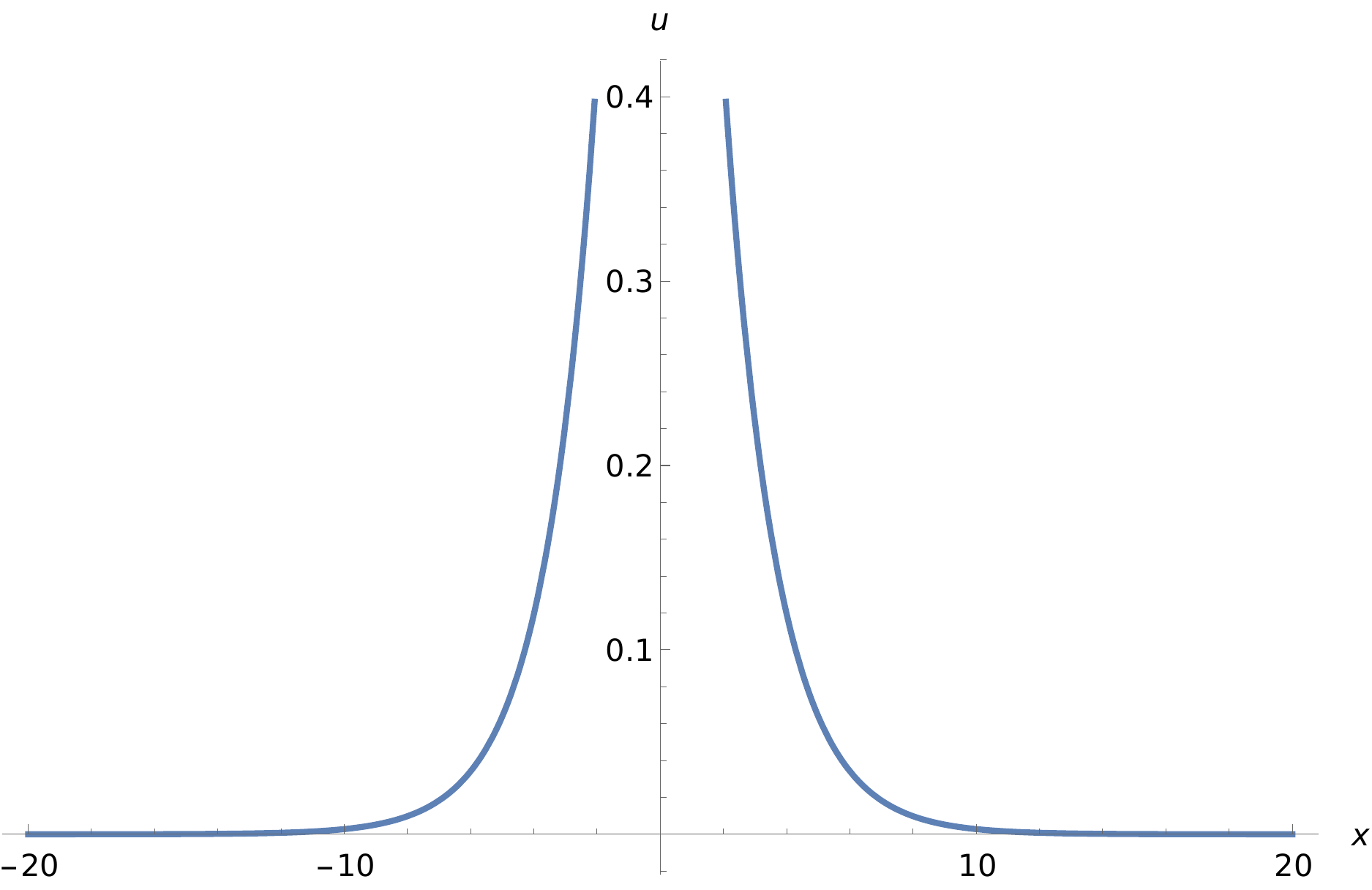}
\hspace{\fill}
\includegraphics[width=0.475\textwidth]{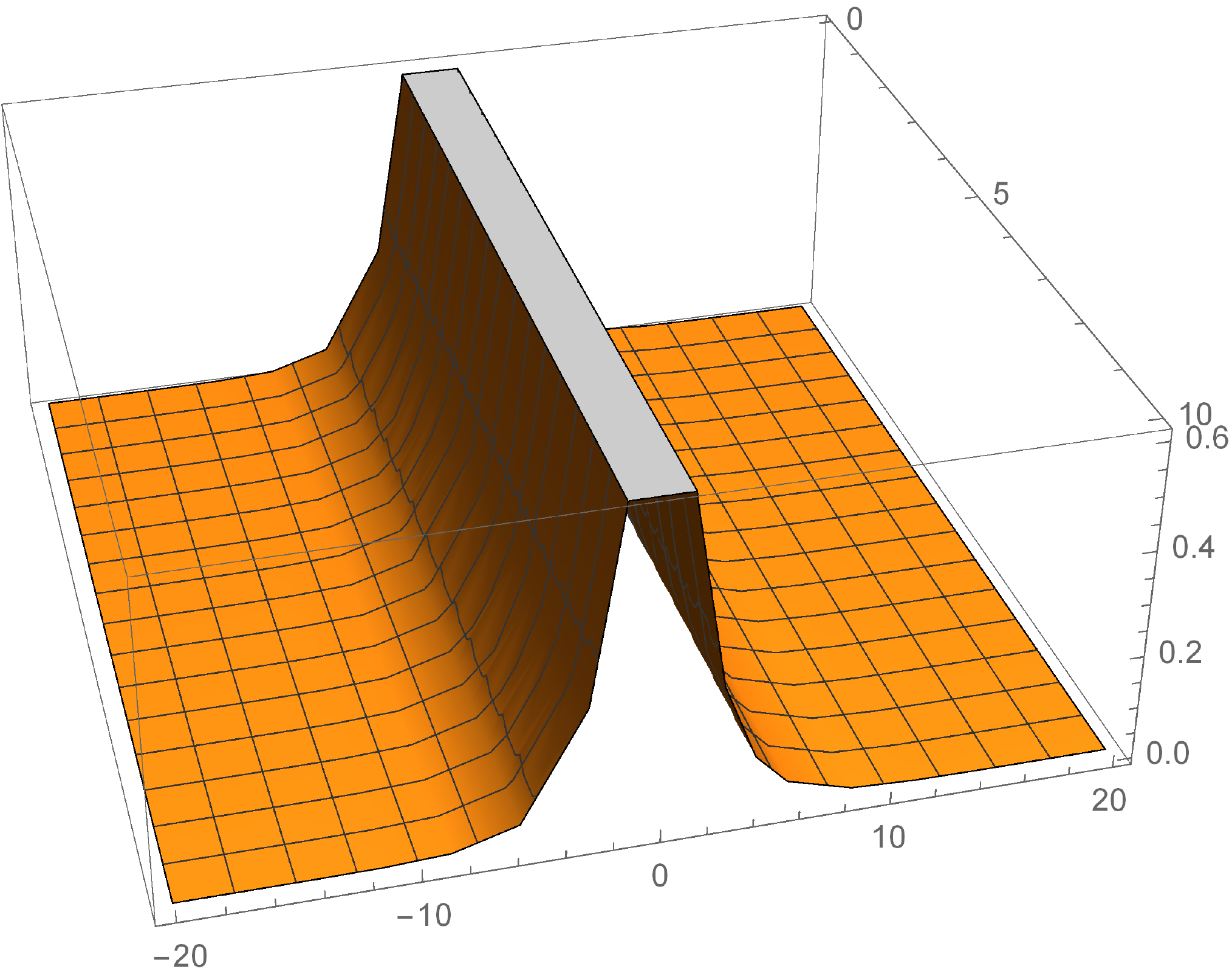}
\caption{2D and 3D shape of  Eq.(\ref{expratcase4}) }\label{fig:13}
\end{figure}

\begin{figure}
\includegraphics[width=0.475\textwidth]{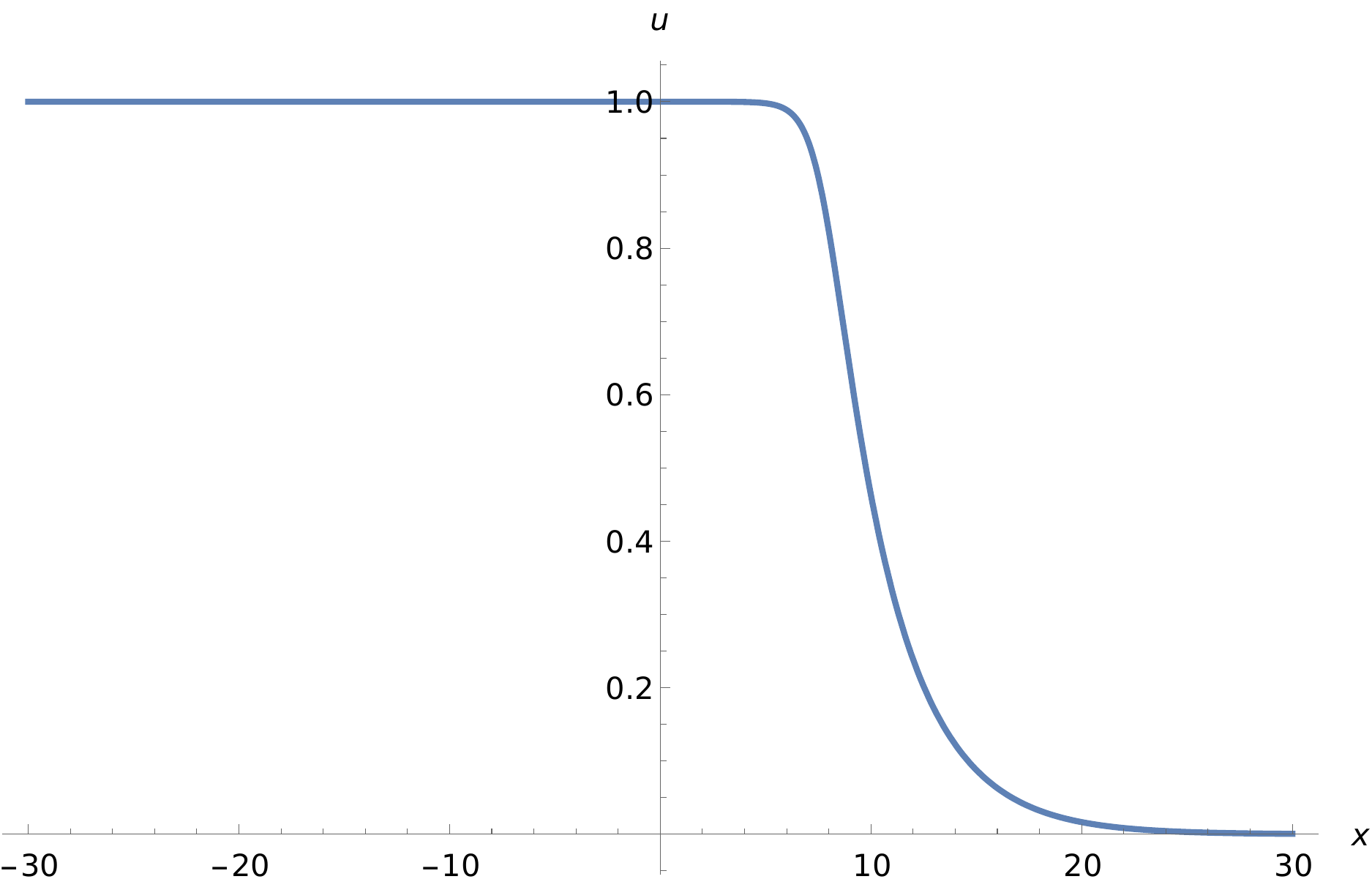}
\hspace{\fill}
\includegraphics[width=0.475\textwidth]{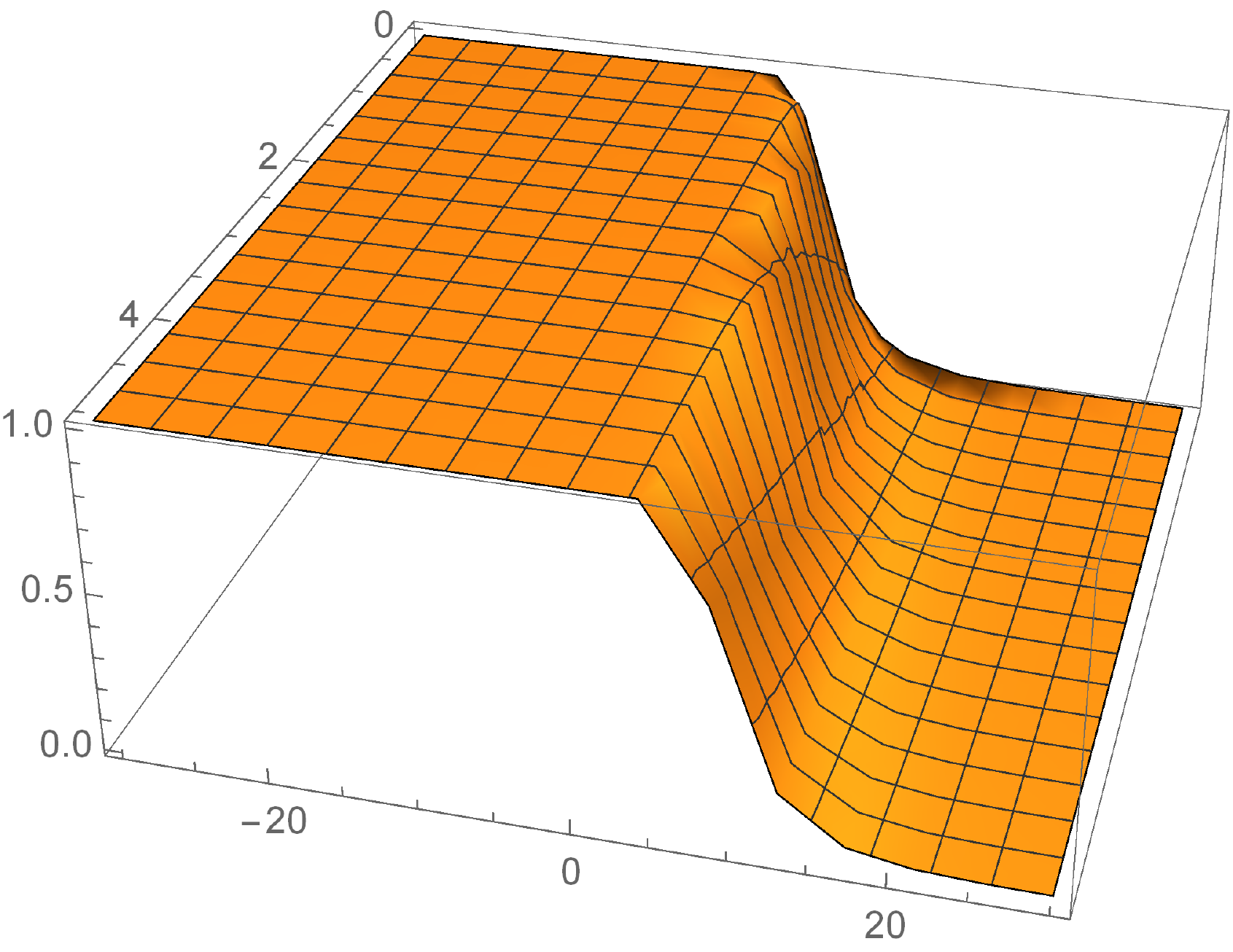}
\caption{2D and 3D shape of  Eq.(\ref{expratcase5}) }\label{fig:14}
\end{figure}

\begin{figure}
\includegraphics[width=0.475\textwidth]{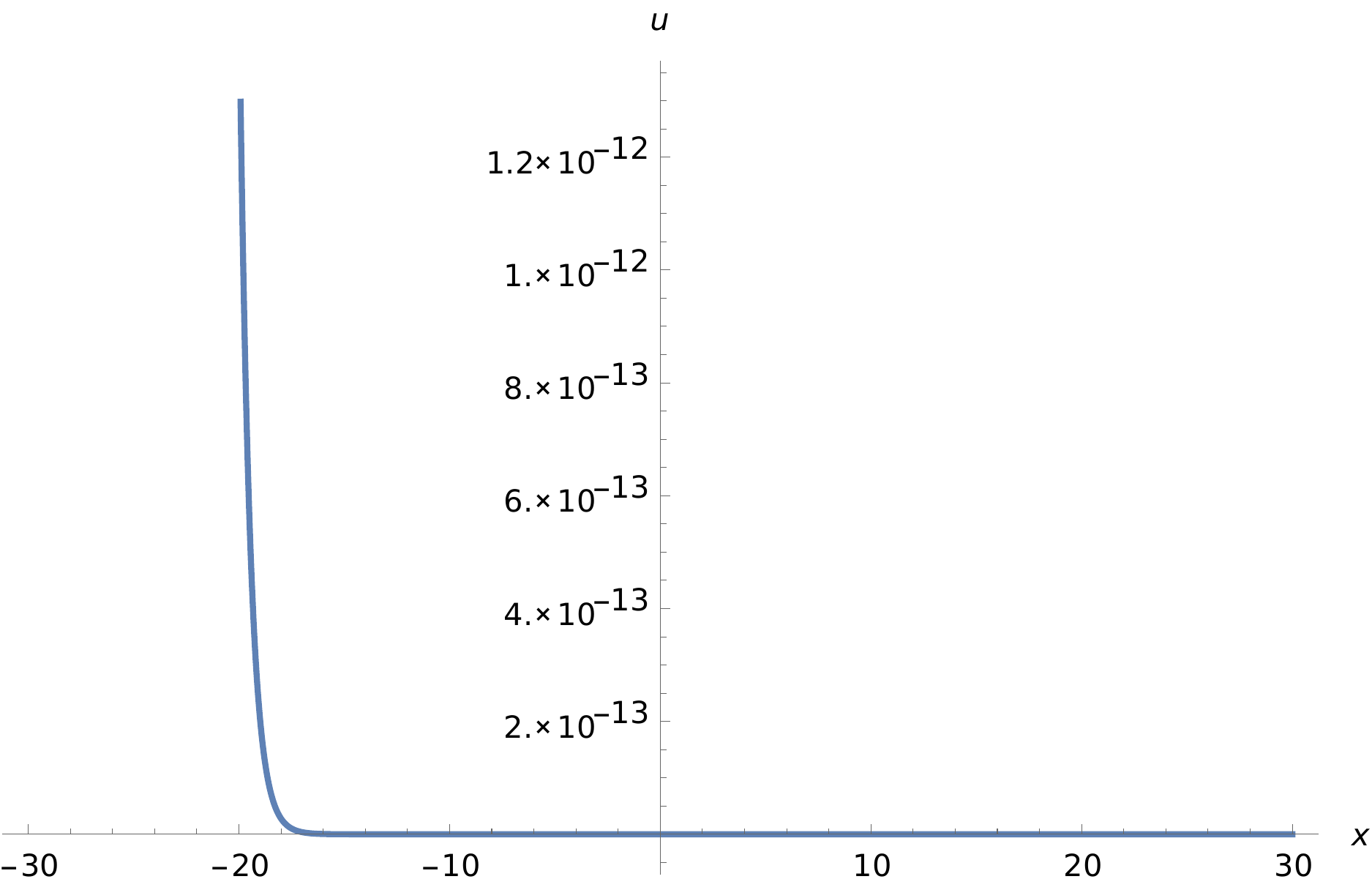}
\hspace{\fill}
\includegraphics[width=0.475\textwidth]{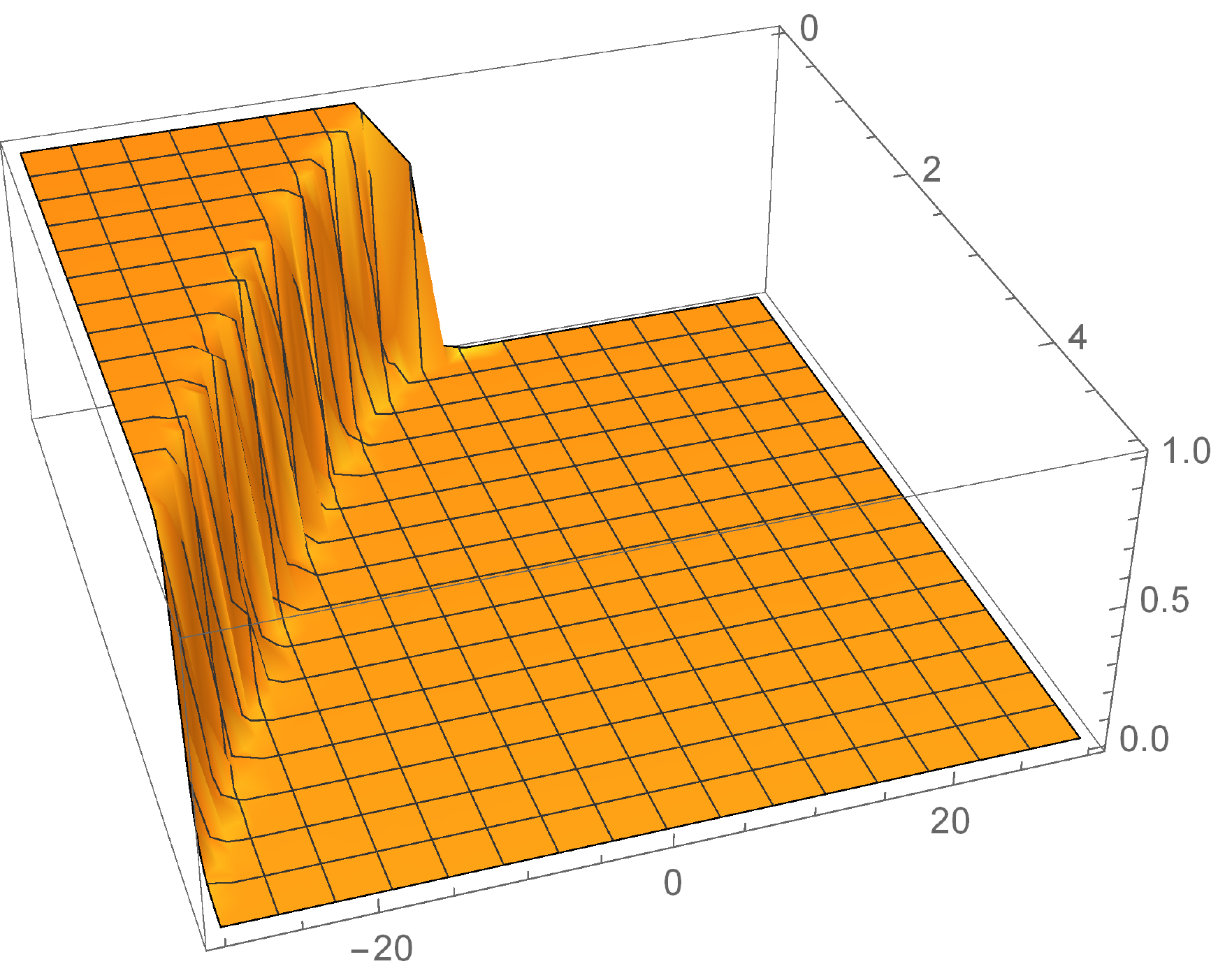}
\caption{2D and 3D shape of  Eq.(\ref{expratcase6}) }\label{fig:15}
\end{figure}

\begin{figure}
\includegraphics[width=0.475\textwidth]{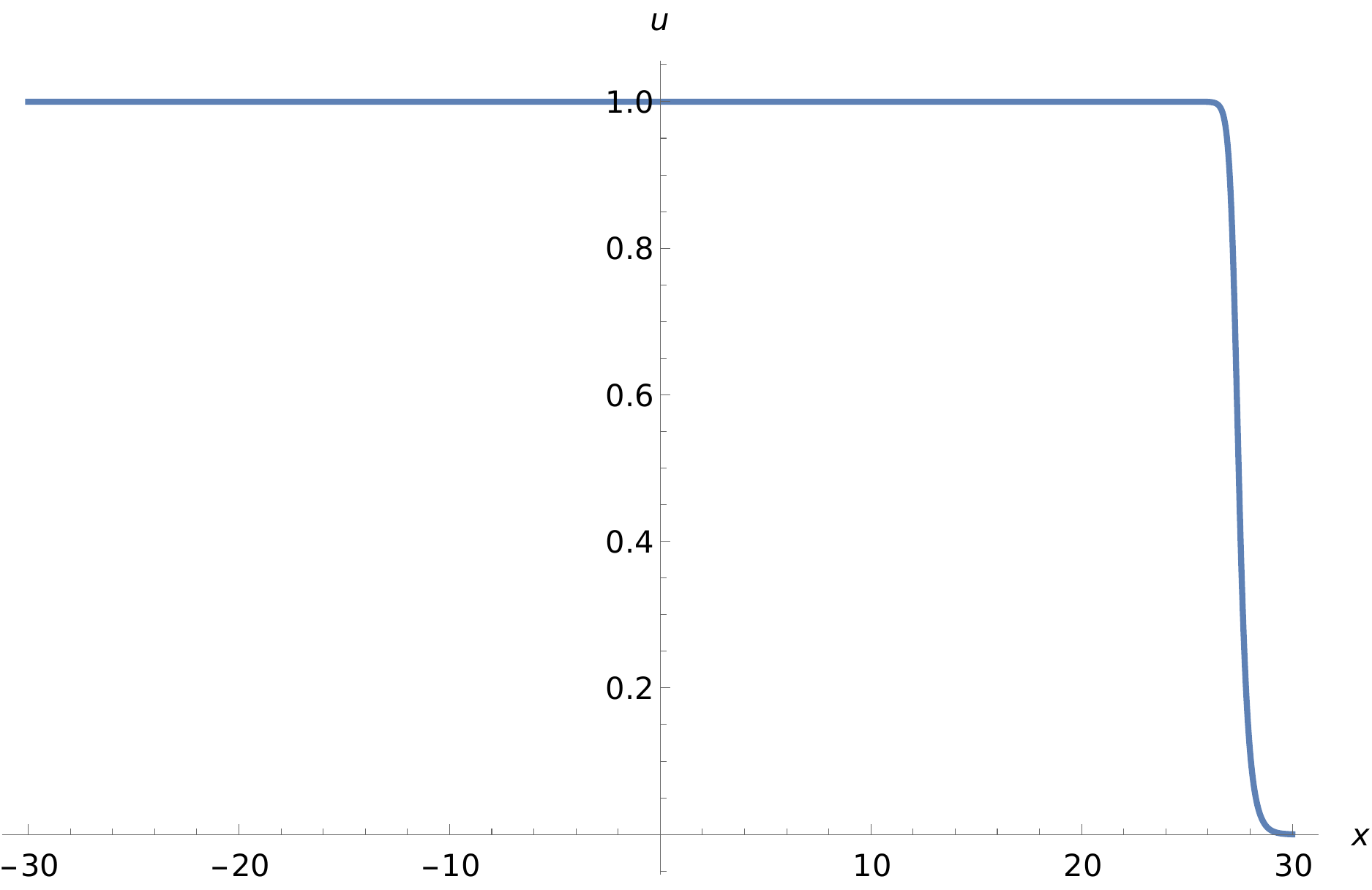}
\hspace{\fill}
\includegraphics[width=0.475\textwidth]{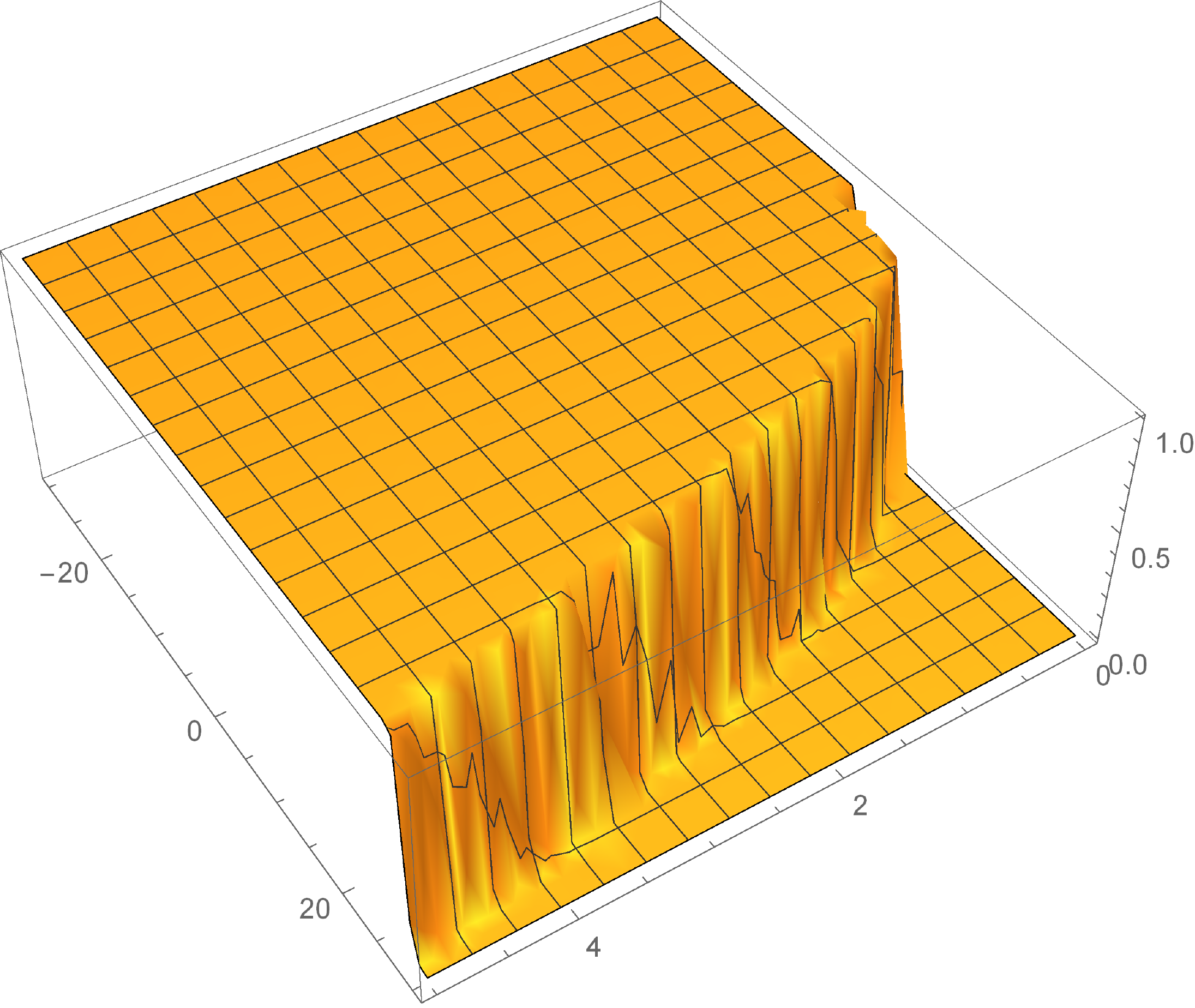}
\caption{2D and 3D shape of  Eq.(\ref{expratcase7}) }\label{fig:16}
\end{figure}

\section{Conclusion}
In this study, the Exp function and exponential rational function methods are used to find the exact solution for the time fractional GBF equation. This generates quite few coefficients and some have been taken into account in providing various analytical solutions. We verified all of the TF-GBF equation's analytic solutions ousing the coefficients and the 2D and 3D plots are also included.

\end{document}